\documentclass[%
 aip,
 jmp,%
 amsmath,amssymb,
 reprint,%
]{revtex4-1}

\usepackage{latexsym}
\usepackage{amsmath}
\usepackage{amsfonts}
\usepackage{natbib}
\usepackage{multirow}
\usepackage{rotating}
\usepackage{graphicx}
\usepackage{dcolumn}
\usepackage{bm}
\usepackage{multirow}
\usepackage{rotating}
\usepackage[table,xcdraw]{xcolor}
\usepackage{graphicx}
\usepackage{gensymb}

\begin{document}


\title[MBE growth of 2\textit{H}-MoTe\textsubscript{2} and 1\textit{T'}-MoTe\textsubscript{2} on 3D substrates]{MBE growth of 2\textit{H}-MoTe\textsubscript{2} and 1\textit{T'}-MoTe\textsubscript{2} on 3D substrates}

\author{Suresh Vishwanath}
\email{sv372@cornell.edu}
\affiliation{School of Electrical and Computer Engineering, Cornell University, Ithaca, NY-14853}
\affiliation{Department of Electrical Engineering, University of Notre Dame, IN-46556}

\author{Aditya Sundar}
\affiliation{Department of Materials Science and Engineering, Cornell University, Ithaca, NY-14853}

\author{Xinyu Liu}
\affiliation{Department of Physics, University of Notre Dame, Notre Dame, IN-46556}

\author{Angelica Azcatl}
\affiliation{Department of Materials Science and Engineering, University of Texas Dallas, Dallas, TX-75083}

\author{Edward Lochocki}
\affiliation{Department of Physics, Laboratory of Atomic and Solid State Physics, Cornell University, Ithaca, New York 14853}

\author{Arthur R. Woll}
\affiliation{Cornell High Energy Synchrotron Source, Cornell University, Ithaca, NY-14853}

\author{Sergei Rouvimov}
\affiliation{Department of Electrical Engineering, University of Notre Dame, IN-46556}

\author{Wan Sik Hwang}
\affiliation{Department of Electrical Engineering, University of Notre Dame, IN-46556}

\author{Ning Lu}
\affiliation{Department of Materials Science and Engineering, University of Texas Dallas, Dallas, TX-75083}

\author{Xin Peng}
\affiliation{Department of Materials Science and Engineering, University of Texas Dallas, Dallas, TX-75083}

\author{Huai-Hsun Lien}
\affiliation{Department of Materials Science and Engineering, Cornell University, Ithaca, NY-14853}

\author{John Weisenberger}
\affiliation{Department of Electrical Engineering, University of Notre Dame, IN-46556} 

\author{Stephen McDonnell}
\affiliation{Department of Materials Science and Engineering, University of Texas Dallas, Dallas, TX-75083}
\affiliation{Department of Materials Science and Engineering, University of Virginia, Charlottesville, VA-22904}

\author{Moon J. Kim}
\affiliation{Department of Materials Science and Engineering, University of Texas Dallas, Dallas, TX-75083}

\author{Margaret Dobrowolska}
\affiliation{Department of Physics, University of Notre Dame, Notre Dame, IN-46556}

\author{Jacek K Furdyna}
\affiliation{Department of Physics, University of Notre Dame, Notre Dame, IN-46556}

\author{Kyle Shen}
\affiliation{Department of Physics, Laboratory of Atomic and Solid State Physics, Cornell University, Ithaca, New York 14853}
\affiliation{Kavli Institute at Cornell for Nanoscale Science, Cornell University, Ithaca, New York 14853}

\author{Robert M. Wallace}
\affiliation{Department of Materials Science and Engineering, University of Texas Dallas, Dallas, TX-75083}

\author{Debdeep Jena}
\affiliation{School of Electrical and Computer Engineering, Cornell University, Ithaca, NY-14853}
\affiliation{Department of Electrical Engineering, University of Notre Dame, IN-46556}
\affiliation{Department of Materials Science and Engineering, Cornell University, Ithaca, NY-14853}

\author{Huili Grace Xing}
\email{grace.xing@cornell.edu}
\affiliation{School of Electrical and Computer Engineering, Cornell University, Ithaca, NY-14853}
\affiliation{Department of Electrical Engineering, University of Notre Dame, IN-46556}
\affiliation{Department of Materials Science and Engineering, Cornell University, Ithaca, NY-14853}
\affiliation{Kavli Institute at Cornell for Nanoscale Science, Cornell University, Ithaca, New York 14853}

\date{\today}

\begin{abstract}
\noindent MoTe\textsubscript{2} is the least explored material in the Molybdenum-chalcogen family, which crystallizes in thermodynamically stable semiconducting 2\textit{H} phase at \textless 500$\degree$C and 1\textit{T'} metallic phase at higher temperatures. Molecular beam epitaxy (MBE) provides an unique opportunity to tackle the small electronegativity difference between Mo and Te while growing layer by layer away from thermodynamic equilibrium. For a few-layer MoTe\textsubscript{2} grown at a moderate rate of  $\sim$6 mins per monolayer under varied Te:Mo flux ratio and substrate temperature, the boundary between the 2 phases in MBE grown MoTe\textsubscript{2} on CaF\textsubscript{2} is characterized using Reflection high-energy electron diffraction (RHEED), Raman spectroscopy and X-ray photoemission spectroscopy (XPS). Grazing incidence X-ray diffraction (GI-XRD) reveals a grain size of $\sim$90 {\AA} and presence of twinned grains. XRD, transmission electron miscroscopy, RHEED, low energy electron diffraction along with lack of electrical conductivity modulation by field effect in MBE 2\textit{H}-MoTe\textsubscript{2} on GaAs (111) B show likelihood of excess Te incorporation in the films. Finally, thermal stability and air sensitivity of MBE 2\textit{H}-MoTe\textsubscript{2} is investigated by temperature dependent XRD and XPS, respectively. 
\end{abstract}

\maketitle

\section{Introduction}

MoTe\textsubscript{2} is still a relatively unexplored transitional metal dichalcogenide (TMD) and holds great promise. MoTe\textsubscript{2} exists in trigonal prismatic (2\textit{H} structure) semiconducting phase at room temperature and metallic 1\textit{T'} phase (monoclinic structure) at high temperatures or metallic \textit{T\textsubscript{d}} phase (orthorhombic structure) when metastable 1\textit{T'} is cooled to $-$33$\degree$C-$-$13$\degree$C\cite{weyl}. 1\textit{T'} and \textit{T\textsubscript{d}} crystal structures have the same in-plane crystal structures but vary in vertical stacking. Monolayer 2\textit{H}-MoTe\textsubscript{2} is predicted to be among the smallest bandgap semiconducting TMDs\cite{KJcho}. Using scanning tunneling spectroscopy (STS) measurements, the bandgap of monolayer 2\textit{H}-MoTe\textsubscript{2} has been measured to be between 1.03 eV\cite{FloridaState}  to 1.4 eV\cite{MoTe2HOPG}. 2\textit{H}-MoTe\textsubscript{2} has a close to broken band alignment with materials like SnSe\textsubscript{2}\cite{Jaggermann}, making them attractive for tunnel based devices such as Esaki diodes\cite{RusenEsaki} and two-dimensional heterojunction interlayer tunneling field effect transistors (Thin-TFETs)\cite{ThinTFET,ThinTFETDRC}. The 1\textit{T'} structure is metastable at room temperature when the bulk MoTe\textsubscript{2} crystal is quenched rapidly. The \textit{T\textsubscript{d}}-MoTe\textsubscript{2} is a type II Weyl semimetal\cite{weyl}, which is a new class of topological material. Traditionally, the transition between 2\textit{H} and 1\textit{T'} phase was thought to be abrupt with respect to temperature\cite{ASM} but recently a mixed phase region has been observed in the phase diagram\cite{MoTe2SDH}. The transition between the 2\textit{H} and 1\textit{T'} phase holds promise for applications such as low resistance contacts\cite{Laseranneal} and phase change memory. 

Large area, phase-pure epitaxial growth with layer controllability would enable industrial applications, but growth of MoTe\textsubscript{2} is especially challenging, since the electronegativity difference between Mo and Te is just 0.3 eV\cite{bernedeXPS,XPSpaper,XPSpaper1}, resulting in a weak bond. Until recently, MoTe\textsubscript{2} has been obtained using tellurization of e-beam deposited molybdenum or molybdenum oxide films\cite{MITMoTe2,KISTMoTe2,zhou2017large} or by chemical vapor transport (CVT)\cite{MoTe2bulkpnjunction,MoTe2SDH}. In case of tellurizing Mo, 1\textit{T'}-MoTe\textsubscript{2} phase is initially formed at 650$\degree$C, which can be converted to 2\textit{H}-MoTe\textsubscript{2} by a 3 hour anneal under Te vapor\cite{KISTMoTe2}. Tellurizing MoO\textsubscript{3} transforms to 2\textit{H}-MoTe\textsubscript{2} more readily, but one cannot be sure in achieving 100\% percent reduction of oxygen. For CVT grown MoTe\textsubscript{2} a mixed phase is observed in the growth temperature range of 500$\degree$C to 900$\degree$C based on the tellurium content in MoTe\textsubscript{x}\cite{MoTe2SDH}, but under ultrahigh vacuum (UHV) in non-thermodynamic equilibrium conditions, this phase boundary is unknown. For example, the formation of a new metallic nanowire phase has been recently reported upon Te loss by annealing at 400-500C in UHV\cite{Mo6Te6}. We recently demonstrated growth of 2\textit{H}-MoTe\textsubscript{2} using molecular beam epitaxy (MBE) in a superlattice with MoSe\textsubscript{2} as well as Bi\textsubscript{2}Te\textsubscript{3}\cite{vishJMR}, where we used a growth temperature of 380$\degree$C. Growth temperature of $\sim$340$\degree$C\cite{UTaustin} or 200$\degree$C\cite{FloridaState,MoTe2HOPG} have been used in other recent MBE demonstrations of 2\textit{H}-MoTe\textsubscript{2} growth. 

In this study, we chose 340$\degree$C to be the lower bound of growth temperature, so as to keep the growth temperature higher than the Te cell temperature ($\sim$300$\degree$C). This avoids intentional accumulation of Te, while keeping the growth temperature significantly lower than the lower bound (500$\degree$C) of the mixed phase, as suggested by the phase diagram under 1 ATM of Te vapor\cite{MoTe2SDH}. We observe that even at a substrate growth temperature as low as 340$\degree$C, the crystalline phase of the MBE-grown MoTe\textsubscript{2} has a sensitive dependence on Te flux in a Te rich environment. Therefore, in Section A of this paper, we present a series of 3 samples of MoTe\textsubscript{2} under different growth conditions and analyze the effect of substrate temperature and tellurium flux on the MoTe\textsubscript{2} crystal structure and stoichiometry. Here, we show that it is indeed feasible to grow phase-pure 2\textit{H}-MoTe\textsubscript{2} on a CaF\textsubscript{2}  substrate without any requirement of a post-growth anneal. In Section B we present the growth and electrical characterization of 2\textit{H}-MoTe\textsubscript{2} on GaAs (111)B. The transition to GaAs was motivated by the availability of high quality epi-ready n\textsuperscript{$+$} GaAs substrates necessary for characterization using techniques such as low energy electron diffraction (LEED). Table.1 summarizes the growth conditions of all the 5 samples in this study. The Mo flux is calculated using the experimentally determined growth rates while assuming zero desorption for molybdenum adatoms, which enables calculation of the Te:Mo flux ratio tabulated in Table.1 for all samples. Growth rate on CaF\textsubscript{2} is calculated using cross-section transmission electron microscopy image and on GaAs (111) B from Reflection high-energy electron diffraction (RHEED) oscillations. 

\begin{sidewaystable}[]
\centering
\caption{Growth conditions for all samples in this study}
\label{my-label}
\begin{tabular}{|c|c|c|c|c|c|c|c|c|}
\hline
Series             & Sample ID & Substrate                                              & \begin{tabular}[c]{@{}c@{}}MoTe\textsubscript{2}\\ Phase\end{tabular} & \begin{tabular}[c]{@{}c@{}}Substrate \\ Temperature ($\degree$C)\end{tabular} & \begin{tabular}[c]{@{}c@{}}Te Flux \\ (Torr)\end{tabular} & \begin{tabular}[c]{@{}c@{}}Te:Mo \\ flux ratio\end{tabular} & \begin{tabular}[c]{@{}c@{}}Growth \\ duration \\ (mins)\end{tabular} & Post growth anneal                                                                       \\ \hline
\multirow{3}{*}{A} & A         & \begin{tabular}[c]{@{}c@{}}CaF\textsubscript{2}\\ (111)\end{tabular}  & 2H                                                    & 340                                                                  & 6.5x10\textsuperscript{-6}                                 & 297                                                          & 30                                                                   & none                                                                                     \\ \cline{2-9} 
                   & B         & \begin{tabular}[c]{@{}c@{}}CaF\textsubscript{2}\\ (111)\end{tabular}   & 2H+1T'                                                & 340                                                                  & 2.0x10\textsuperscript{-6}                                 & 98                                                          & 30                                                                   & none                                                                                     \\ \cline{2-9} 
                   & C         & \begin{tabular}[c]{@{}c@{}}CaF\textsubscript{2}\\ (111)\end{tabular}   & 1T'                                                   & 400                                                                  & 1.4x10\textsuperscript{-6}                                 & 71                                                         & 30                                                                   & none                                                                                     \\ \hline
\multirow{2}{*}{B} & D         & \begin{tabular}[c]{@{}c@{}}GaAs\\ (111) B\end{tabular} & 2H                                                    & 340                                                                  & 6.9x10\textsuperscript{-6}                                 & 175                                                         & 30                                                                   & \begin{tabular}[c]{@{}c@{}}at 380$\degree$C for 10 mins \\ without Te\end{tabular}                \\ \cline{2-9} 
                   & E         & \begin{tabular}[c]{@{}c@{}}GaAs\\ (111) B\end{tabular} & 2H                                                    & 340                                                                  & 6.9x10\textsuperscript{-6}                                 & 262                                                         & 20                                                                   & \begin{tabular}[c]{@{}c@{}}at 450$\degree$C for 3 mins and \\ 550$\degree$C for 7 mins under Te\end{tabular} \\ \hline
\end{tabular}
\end{sidewaystable}
 
\section{Section A : M\MakeLowercase{o}T\MakeLowercase{e}\textsubscript{2} \MakeLowercase{on} C\MakeLowercase{a}F\textsubscript{2}}

\subsection{Growth conditions}

Three samples (see Table.1) constitute the series of samples grown on CaF\textsubscript{2}. Calcium fluoride was chosen as a substrate because (i) it has an inert fluorine-terminated surface on which we have successfully grown MoSe\textsubscript{2}\cite{vish2D} and (ii) it provides a cavity effect enhancing the Raman signal due to its optical transparency. On the other hand, the Raman signal from MoTe\textsubscript{2} grown on GaAs(111)B is very weak\cite{vishJMR}. These samples were grown in a Riber 32 MBE system using elemental Mo delivered from an e-beam evaporator and elemental uncracked Te from a Knudsen cell. The growth duration was 30 mins for each sample and the Mo flux, which limits the growth rate, was set to $\sim$0.17 monolayer (ML) per minute or 6 minutes per ML. Supplementary (SI) Fig.1 shows that all CaF\textsubscript{2} substrates were first heated to 800$\degree$C, held for 30 mins in order to anneal and degas. Sharp RHEED streaks of CaF\textsubscript{2} prior to start of growth (see Fig.\ref{RHEED}(a)), show the smooth crystalline post-anneal growth surface. Then the substrates were lowered to respective growth temperatures, stabilized for $\sim$30 mins prior to thin film growth. The growth conditions are listed in Table.1. Using these 3 samples, we observe that, although the Te flux is 2 orders of magnitude higher than the Mo flux, the Te flux range to obtain 2\textit{H}-MoTe\textsubscript{2} is narrow and the substrate temperature control is critical. None of the MoTe\textsubscript{2} films on CaF\textsubscript{2} have been annealed in order to avoid phase change during annealing. All temperatures given in this study are thermocouple temperatures and the sample surface temperature is estimated about 20$\degree$C lower than the thermocouple temperature.

\subsection{Results and discussion}

\subsubsection{Reflection high-energy electron diffraction (RHEED)}

\begin{figure}
	\includegraphics[width=\columnwidth]{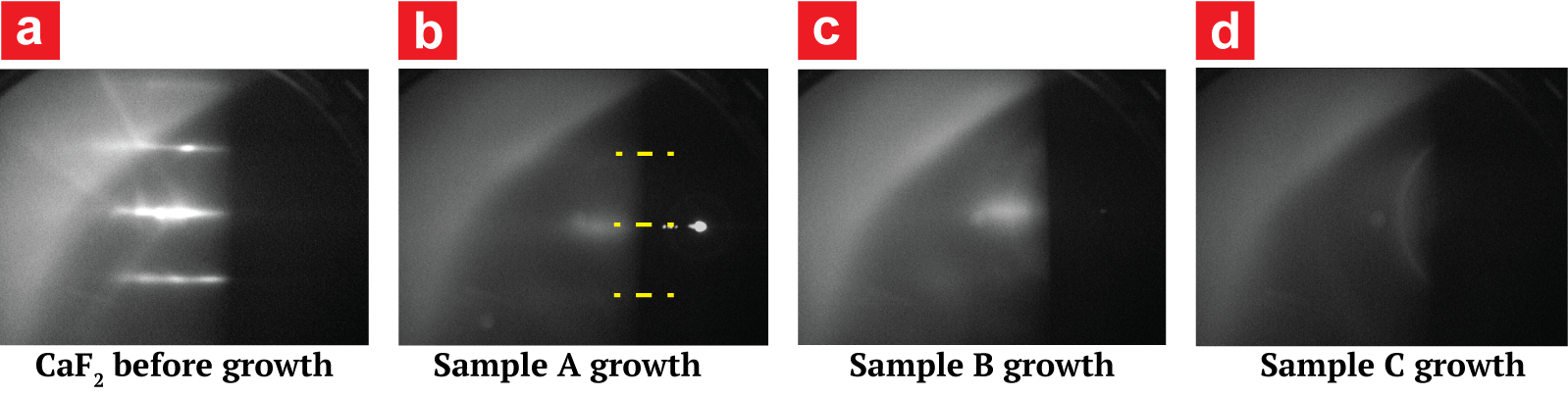}
	\caption {{(a-d)RHEED pattern from the series of samples on CaF\textsubscript{2} showing that 1\textit{T'} phase grows polycrystalline as compared to well-aligned growth of 2\textit{H} phase.} \label{RHEED}}
\end{figure}

As seen in Fig.\ref{RHEED}(c), annealed CaF\textsubscript{2} has a strong streaky RHEED pattern prior to growth. Sample A shows a more diffused but still streaky RHEED pattern, whereas sample B shows a mixed pattern comprising of streaks and a ring, which evolves into only a ring in sample C. Streaky RHEED points to layered growth with minimal mosaicity of the as-grown film, but progressive inclusion of a ring points to another growth mechanism taking over, which results in polycrystalline growth. Whether the polycrystalline material is the same phase as the streaky film or a different phase is elucidated through employing Raman and XPS characterization (described below). The RHEED streaks of MoTe\textsubscript{2} in sample A along \textless11$\bar{2}$0\textgreater\ appear at the same position as the \textless1$\bar{1}$0\textgreater\  of CaF\textsubscript{2}, as observed previously in MoSe\textsubscript{2} on CaF\textsubscript{2}\cite{vish2D,koma}. The lattice spacing of MoTe\textsubscript{2} based upon the ratio of the RHEED streak spacing is $\sim$3.5 {\AA}, which is very close to the value of 3.52 {\AA}\cite{MoTe2SDH} corresponding to bulk 2\textit{H}-MoTe\textsubscript{2}. The ambiguity in this measurement is due to the diffused RHEED pattern, which is measured more accurately using grazing incidence X-ray diffraction (GI-XRD), presented further below.

\subsubsection{Raman Spectroscopy}

\begin{table}[]
\centering
\caption{Positions of the various Raman peaks compared to measured values for bulk 2\textit{H}-MoTe\textsubscript{2} and reported values for 1\textit{T'}-MoTe\textsubscript{2}\cite{KISTMoTe2}.}
\label{my-label}
\begin{tabular}{|c|c|c|c|c|}
\hline
\begin{tabular}[c]{@{}c@{}}Sample\\ ID\end{tabular} & \begin{tabular}[c]{@{}c@{}}A\textsubscript{1g} position\\ (cm\textsuperscript{-1})\end{tabular} & \begin{tabular}[c]{@{}c@{}}E\textsuperscript{2}\textsubscript{2g} position\\ (cm\textsuperscript{-1})\end{tabular} & \begin{tabular}[c]{@{}c@{}}B\textsubscript{g} position\\ (cm\textsuperscript{-1})\end{tabular} & \begin{tabular}[c]{@{}c@{}}A\textsubscript{g} position\\ (cm\textsuperscript{-1})\end{tabular} \\ \hline
Bulk 2\textit{H}                                             & 174.47                                                        & 235.80                                                         & NA                                                           & NA                                                           \\ \hline
A                                                   & 173.30                                                        & 235.79                                                         & NA                                                           & NA                                                           \\ \hline
B                                                   & 174.14                                                        & 236.66                                                         & 158.07                                                       & 255.59                                                       \\ \hline
C                                                   & NA                                                            & NA                                                             & 159.14                                                       & 256.23                                                       \\ \hline
1\textit{T'}\cite{KISTMoTe2}                                                 & NA                                                            & NA                                                             & 163                                                          & 256.1                                                        \\ \hline
\end{tabular}
\end{table}

\begin{figure}
	\includegraphics[scale=0.25]{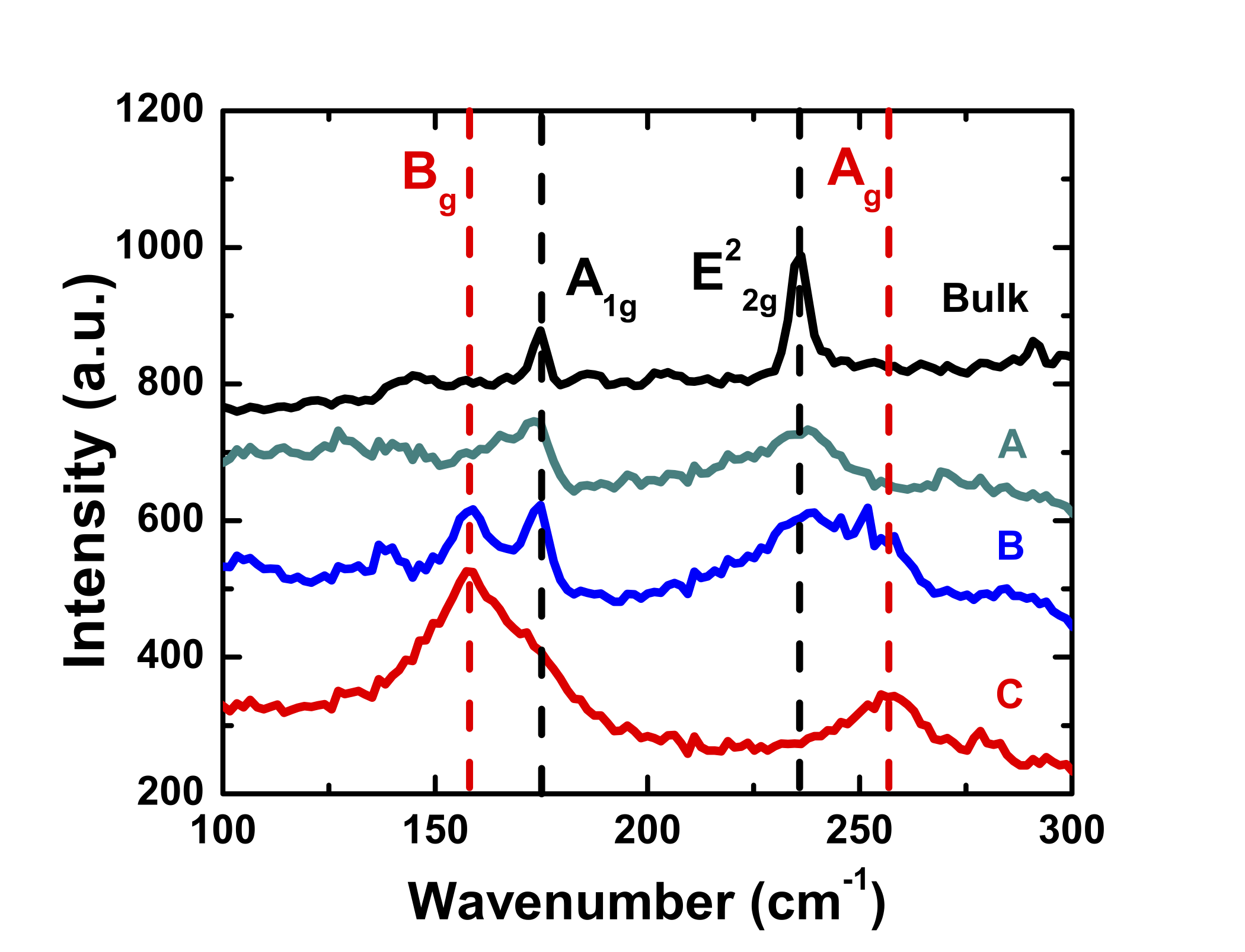}
	\caption{(a) Raman measurements on the samples grown on CaF\textsubscript{2}. \label{raman}}
\end{figure}

Raman spectra in Fig.\ref{raman} confirm an evolution from the 2\textit{H} phase to a new phase as we progress from sample A to sample C. The 2\textit{H} phase is confirmed by comparing Raman from sample A with Raman from CVT grown bulk 2\textit{H}-MoTe\textsubscript{2} obtained from 2D Semiconductors Inc. as shown in Table.2. Its important to note that the FWHM of the peaks from MBE grown MoTe\textsubscript{2} is several times wider than that of the CVT-grown MoTe\textsubscript{2}. This points to a significant disorder in the MBE-grown material and augments the observation of the diffuse MoTe\textsubscript{2} RHEED pattern in sample A. The new phase is assigned to the 1\textit{T'} phase due to its proximity to the peak position reported for 1\textit{T'} in literature\cite{KISTMoTe2}. So far, only 2\textit{H} and 1\textit{T'} or \textit{Td} phases of MoTe\textsubscript{2} are known. As seen in Table.2 the A\textsubscript{g} peak position of the 1\textit{T'} phase at 255.59 cm\textsuperscript{-1}-256.23 cm\textsuperscript{-1} agrees closely with the reported value of 256.1 cm\textsuperscript{-1} - 257 cm\textsuperscript{-1} for 1\textit{T'} MoTe\textsubscript{2} growth by tellurization of molybdenum films but the B\textsubscript{g} peak at 158.07 cm\textsuperscript{-1}-159.14 cm\textsuperscript{-1} deviates significantly from the reported value of 163 cm\textsuperscript{-1} - 161 cm\textsuperscript{-1} and extremely broad\cite{KISTMoTe2}\cite{MITMoTe2} . Sample B shows a mixed phase comprised of Raman signatures from both phases, where as sample C only shows 1\textit{T'} peaks.

\subsubsection{X-ray photoemission spectroscopy (XPS)}

\begin{sidewaystable}[]
\small
\centering
\caption{XPS peak positions for Mo 3d\textsubscript{5/2} and Te 3d\textsubscript{5/2} as well as Te:Mo ratio corresponding to both phases 2\textit{H}-MoTe\textsubscript{2} and 1\textit{T'}-MoTe\textsubscript{2} compared to reported values}
\label{my-label}
\begin{tabular}{|c|c|c|c|c|c|c|c|c|c|}
\hline
\begin{tabular}[c]{@{}c@{}}Sample \\ ID\end{tabular}              & \begin{tabular}[c]{@{}c@{}}Te:Mo \\ flux\\ ratio\end{tabular} & \begin{tabular}[c]{@{}c@{}}Mo 3d\textsubscript{5/2}\\ 2\textit{H} \\ (eV)\end{tabular} & \begin{tabular}[c]{@{}c@{}}Te 3d\textsubscript{5/2}\\ 2\textit{H} \\ (eV)\end{tabular} & \begin{tabular}[c]{@{}c@{}}Mo 3d\textsubscript{5/2}\\ 1\textit{T'} \\ (eV)\end{tabular} & \begin{tabular}[c]{@{}c@{}}Te 3d\textsubscript{5/2}\\ 1\textit{T'} \\ (eV)\end{tabular} & \begin{tabular}[c]{@{}c@{}}Te:Mo\textsubscript{2\textit{H}} \\ ratio\\ w/o corr.\end{tabular} & \begin{tabular}[c]{@{}c@{}}Te:Mo\textsubscript{2\textit{H}}\\ ratio\\ corr.\end{tabular} & \begin{tabular}[c]{@{}c@{}}Te:Mo\textsubscript{1\textit{T'}}\\ ratio\\ w/o corr.\end{tabular} & \begin{tabular}[c]{@{}c@{}}Te:Mo\textsubscript{1\textit{T'}}\\ ratio\\ corr.\end{tabular} \\ \hline
\begin{tabular}[c]{@{}c@{}}Bulk 2\textit{H}\\ \end{tabular} &                                                               & \begin{tabular}[c]{@{}c@{}}227.8\\ (air exposed\cite{XPSpaper})\\228.0 \\(as cleaved\cite{XPSpaper})\\227.8 \\(as cleaved\cite{bernedeXPS})\end{tabular}                                                        & \begin{tabular}[c]{@{}c@{}}572.9\\ (air exposed\cite{XPSpaper})\\573.0 \\(as cleaved\cite{XPSpaper}) \\572.4 \\(as cleaved\cite{bernedeXPS})\end{tabular}                                                         &                                                                &                                                                &                                                                     & \begin{tabular}[c]{@{}c@{}}\\ \\ \\ \\ \\ 2.03\end{tabular}                                                                 &                                                                    &                                                                  \\ \hline
A                                                                 & 297                                                            & 227.9                                                         & 572.8                                                         &                                                                &                                                                &                                                                     & 2.57                                                             &                                                                    &                                                                  \\ \hline
B                                                                 & 98                                                            & 227.9                                                         & 572.9                                                         & 227.4                                                          & 572.6                                                          & 2.03                                                                & 2.54                                                             & 2.12                                                               & 2.66                                                             \\ \hline
C                                                                 & 71                                                           & 228                                                           & 572.9                                                         & 227.5                                                          & 572.6                                                          & 2.04                                                                & 2.61                                                             & 2.13                                                               & 2.72                                                             \\ \hline
1\textit{T'}                                                               &                                                               &                                                               &                                                               & \begin{tabular}[c]{@{}c@{}}227.7\cite{naylorUpenn}\\ 228\cite{MITMoTe2}\end{tabular}            & \begin{tabular}[c]{@{}c@{}}572.1\cite{naylorUpenn}\\ 572.6\cite{MITMoTe2}\end{tabular}          &                                                                     &                                                                  &                                                                    &                                                                  
\\ \hline
\end{tabular}
\end{sidewaystable}

\begin{figure}
	\includegraphics[width=\columnwidth]{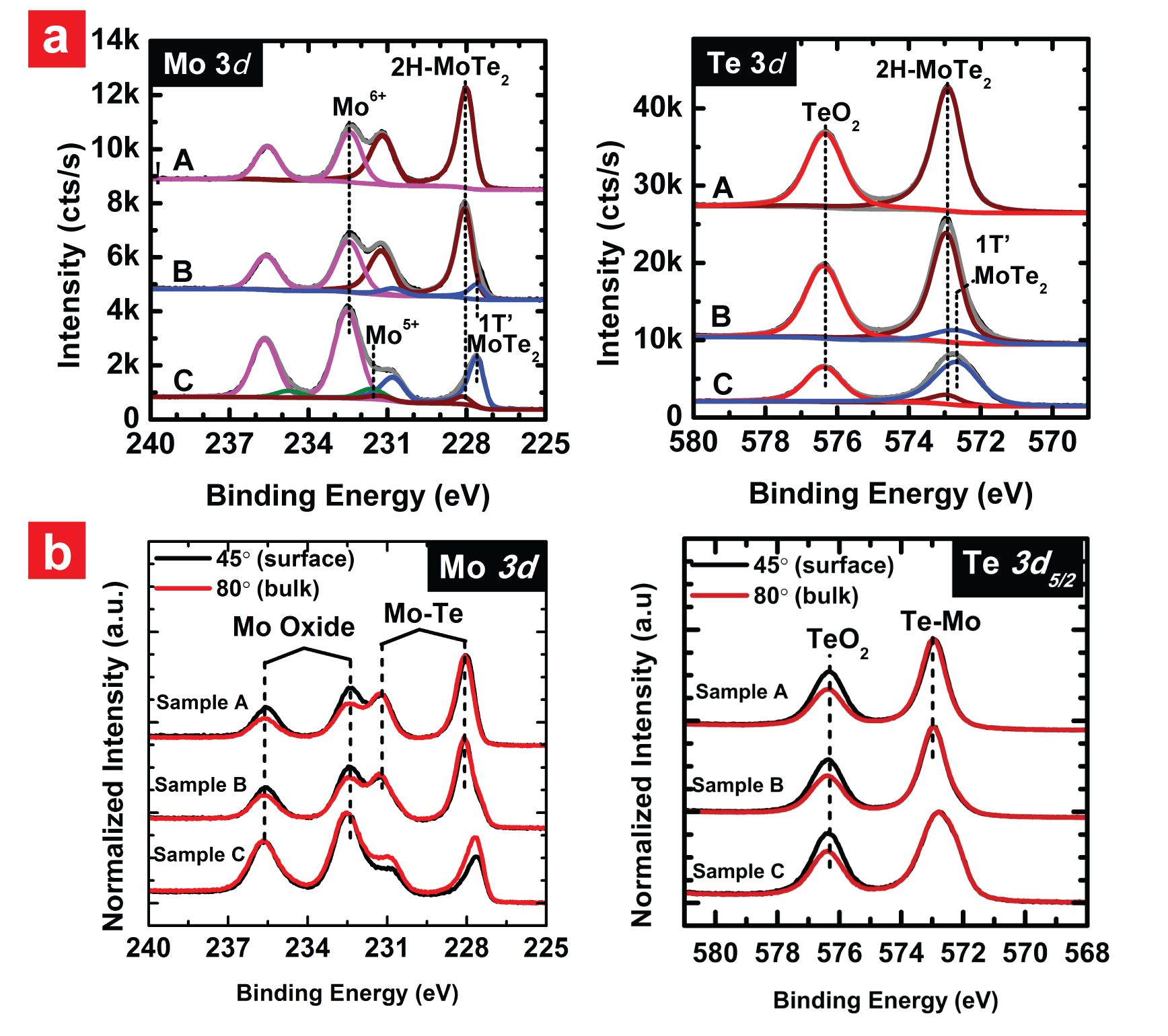}
	\caption{(a) XPS on samples showing various phases and the extent of oxidation under 
		various growth conditions. The pink line corresponds to Mo\textsuperscript{+6} oxide, maroon line to 2\textit{H}-MoTe\textsubscript{2}, blue line to 1\textit{T'}-MoTe\textsubscript{2}, red line to TeO\textsubscript{2} and green line to Mo\textsuperscript{+5} oxide (b) shows that the majority of the oxidation is limited to the surface by comparing the oxide signal from the surface and the bulk by varying the take-off angle.\label{XPS}}
\end{figure}

XPS spectra corresponding to Mo, Te, O, Ca, F and C are detected from all samples (Fig.\ref{XPS}). No charging effects were detected on any of them. Peak positions for Mo 3d\textsubscript{5/2} and Te 3d\textsubscript{5/2} as well as Te:Mo ratio corresponding to both phases 2\textit{H}-MoTe\textsubscript{2} and 1\textit{T'}-MoTe\textsubscript{2} are listed in Table.3. For sample A, the Mo 3d\textsubscript{5/2} signal corresponding to 2\textit{H}-MoTe\textsubscript{2} bond was detected at 227.9 eV, which is consistent with the binding energy of 2\textit{H}-MoTe\textsubscript{2} in literature\cite{XPSpaper}. The Te 3d\textsubscript{5/2} peak corresponding to the 2\textit{H}-MoTe\textsubscript{2} is observed at 572.8 eV. Molybdenum oxide in the Mo\textsuperscript{+6} state was also identified. In the Te 3d spectrum, tellurium dioxide and MoTe\textsubscript{2} are both detected. The Te:Mo ratio is $\sim$2.57 after correction due to attenuation from the oxide overlayer. For sample B, in addition to the peaks corresponding to 2\textit{H}-MoTe\textsubscript{2}, Mo\textsuperscript{+6} oxide and TeO\textsubscript{2} are detected; the Mo 3d\textsubscript{5/2} peak at 227.6 eV and the Te 3d\textsubscript{5/2} peak at 572.6 eV are assigned to the 1\textit{T'} phase. The Te:Mo ratio for the 2\textit{H}-MoTe\textsubscript{2} component is calculated to be 2.03. After correction due to attenuation from the oxide overlayer the Te:Mo ratio is calculated to be 2.54. The Te:Mo ratio for the 1\textit{T'}-MoTe\textsubscript{2} component in sample B is 2.12 and, when corrected for the oxide overlayer is 2.66. For sample C, there is a very small signal from 2\textit{H}-MoTe\textsubscript{2} with Mo 3d\textsubscript{5/2} at 228 eV and the corresponding peak for Te 3d\textsubscript{5/2} at 572.7 eV. But the majority of the MoTe\textsubscript{2} peak intensity is from a new Mo 3d\textsubscript{5/2} peak at 227.5 eV and Te 3d\textsubscript{5/2} at 572.5 eV, which are assigned to chemical states associated with the 1\textit{T'} phase of MoTe\textsubscript{2}. The Te:Mo ratio corresponding to 1\textit{T'} phase of MoTe\textsubscript{2} is 2.13 and, when corrected for the oxide overlayer is 2.72. The Te:Mo ratio for the 2\textit{H}-MoTe\textsubscript{2} component is 2.04 and, when corrected for the oxide overlayer, is 2.61. It is key to note that in sample C the Mo 3d peak intensity associated with Mo oxide is much higher than that for MoTe\textsubscript{2}, as well as the oxide intensity from the other samples. This suggests that, inspite of employing a large over pressure of uncracked Te (dimers) during growth, not only does Mo form predominantly 1\textit{T'}-MoTe\textsubscript{2} but that majority of Mo has an increased propensity for oxidation. This molybdenum oxide in sample C exhibits 2 different Mo oxidation states of +5 and +6. Reported peak position for 1\textit{T'}-MoTe\textsubscript{2} for Mo 3d\textsubscript{5/2} is 227.7-228 eV and for Te 3d\textsubscript{5/2} is 572.1-572.6 eV\cite{naylorUpenn,MITMoTe2}.  The observed XPS peak from the phase assigned to 1\textit{T'}-MoTe\textsubscript{2} for Te 3d\textsubscript{5/2} is consistent with the reported value but that for Mo 3d\textsubscript{5/2} is much lower that what has been reported for any Mo-Te bond and even metallic Mo 3d\textsubscript{5/2} at 227.8 eV\cite{werfel1983photoemission}. It is also noted that in all 3 samples, the oxide peaks from Mo and Te in the O1s spectra could not be resolved because of the close proximity in electronegativity of Mo and Te\cite{bernedeXPS}. The O 1s spectral feature also has contributions from C-O and O-H. Fig.\ref{XPS}(b) shows that concentration of both molybdenum oxide and tellurium oxide concentrations are higher on the surface as seen from the increase in intensity of oxides at take-off angle of 45$\degree$ as compared to 80$\degree$. (A take-off angle of 80$\degree$ is much more bulk sensitive than 45$\degree$.) Also, the chemical bonding state of MoTe\textsubscript{2} is homogeneous through the analyzed depth, as the Mo-Te peak width remains constant when changing angle. Its worthy to note that the extent of oxidation in the telluride system is much more than previously reported MBE MoSe\textsubscript{2}\cite{vish2D}. 

\subsubsection{Grazing incidence X-ray diffraction (GI-XRD) and Transmission electron microscopy (TEM)}

\begin{figure}
	\includegraphics[width=0.8\columnwidth]{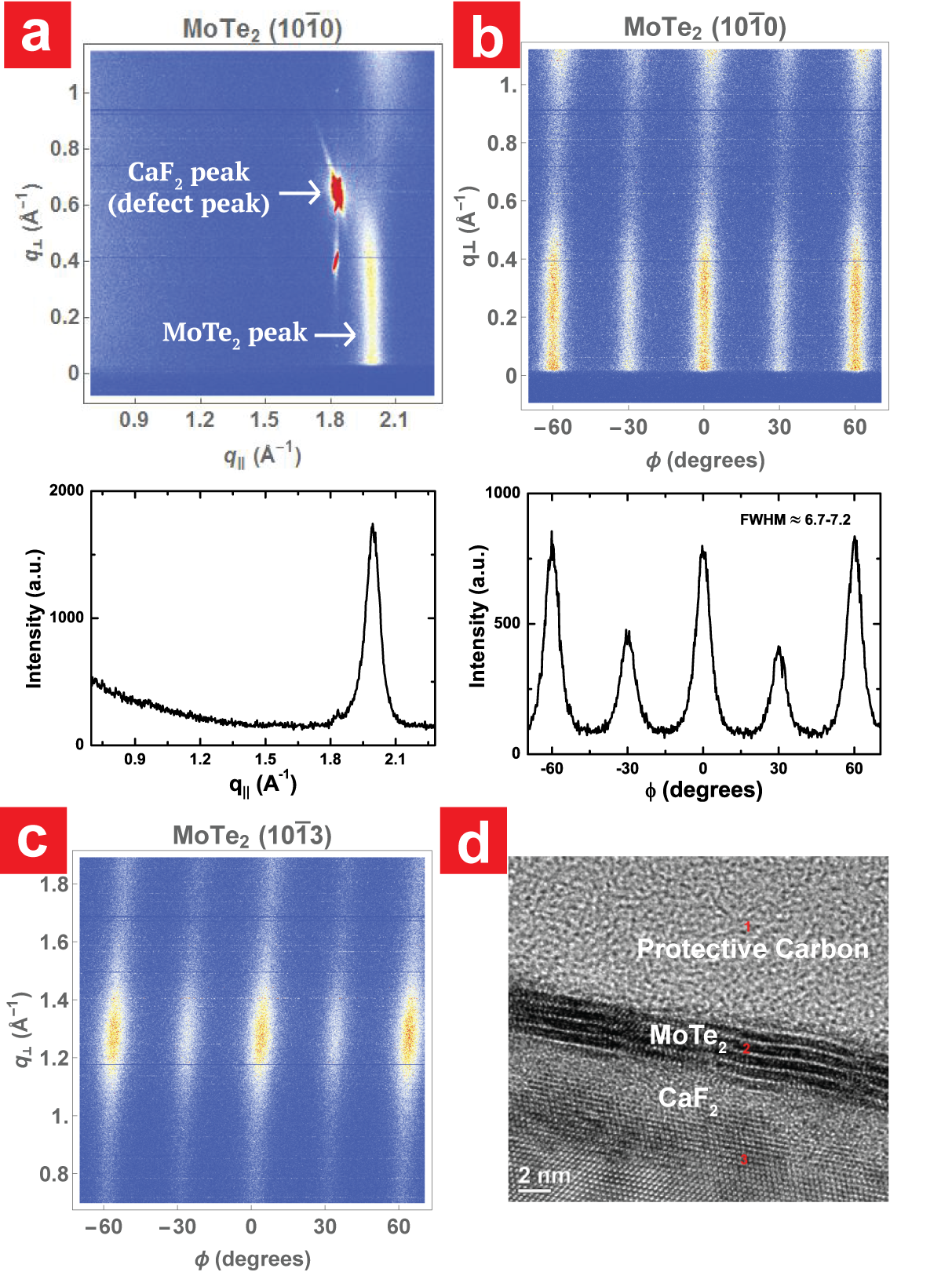}
	\caption{For sample A (a) in-plane $\omega$-2$\theta$ scan showing a peak corresponding to CaF\textsubscript{2} and 
		$\{$10$\bar{1}$0$\}$ of MoTe\textsubscript{2}. The extended streak in the perpendicular direction is due to overlap of the extended rods from $\{$10$\bar{1}$0$\}$ and $\{$10$\bar{1}$1$\}$ in reciprocal space due to the $\sim$5 monolayer thin film. Below it is the integrated intensity in a range of 0.02 to 0.15 
		{\AA}\textsuperscript{-1}	q\textsubscript{$\perp$} corresponding to $\{$10$\bar{1}$0$\}$ peak(b) The in-plane phi($\phi$) scan of the $\{$10$\bar{1}$0$\}$ peak of MoTe\textsubscript{2} to understand the rotational alignment and the extent of twinning in the grown film. Below it is the integrated intensity in a range of 0.02 to 0.15 
		{\AA}\textsuperscript{-1}	q\textsubscript{$\perp$} (c) It shows the periodicity of  $\{$10$\bar{1}$3$\}$ peak. (d) cs-TEM.\label{GIXRD}}
\end{figure}

In order to get a better estimation of the inplane lattice constant as compared to the estimation using RHEED pattern, and to understand the preference of in-plane rotational orientation, grazing incidence X-ray diffraction (GI-XRD) was done. GI-XRD from sample A shows (see Fig.\ref{GIXRD}(a)) an extended line corresponding to overlapped $\{$10$\bar{1}$0$\}$ and $\{$10$\bar{1}$1$\}$ set of planes of MoTe\textsubscript{2} and the sharp high intensity peak is from the CaF\textsubscript{2} substrate. The in-plane lattice constant of 2\textit{H}-MoTe\textsubscript{2} calculated from the $\{$10$\bar{1}$0$\}$ peak corresponds to 3.638 {\AA}. The full width half maximum (FWHM) of 2\textit{H}-MoTe\textsubscript{2} $\{$10$\bar{1}$0$\}$ peak is calculated to be $\sim$0.079 {\AA}\textsuperscript{-1} and the direct beam FWHM is $\sim$0.009 {\AA}\textsuperscript{-1}. By subtracting the width of the direct beam, the genuine FWHM of the $\{$10$\bar{1}$0$\}$ peak was estimated to be $\sim$0.07 {\AA}\textsuperscript{-1} which translates to a grain size of $\sim$92 {\AA}\cite{GIXRD_1}. From Fig.\ref{GIXRD}(b), which is an inplane phi($\phi$) scan, we can see that MBE MoTe\textsubscript{2} undergoes significant twinning thus showing 2 sets of 6-fold symmetry diffraction patterns. The peak ratio between adjacent peaks separated by 30$\degree$  is $\sim$0.5. This shows that almost 30\% of the grains are twins. Also the wide FWHM (6.7$\degree$-7.2$\degree$) of these peaks signifies a large deviation of grains from the preferred orientation. From measurements and simulation it has been shown in 2\textit{H}-MoSe\textsubscript{2} that $\{$10$\bar{1}$1$\}$ and $\{$10$\bar{1}$2$\}$ peaks are $\sim$10 times weaker than $\{$10$\bar{1}$3$\}$\cite{MoSe2XRD}. 2\textit{H}-MoTe\textsubscript{2} having the same crystal structure as 2\textit{H}-MoSe\textsubscript{2}, we also observe the $\{$10$\bar{1}$3$\}$ set of peaks at higher q\textsubscript{$\perp$} (see Fig.\ref{GIXRD}(c)), from which the out of plane lattice constant (c-spacing) is calculated to be 14.4 {\AA}. The c axis lattice constant obtained from cs-TEM as shown in Fig.\ref{GIXRD}(d) is 13.9 {\AA}. The reported value for the inplane lattice constant and c axis lattice constant from bulk 2\textit{H}-MoTe\textsubscript{2} are 3.52 {\AA} and 13.966 {\AA}, respectively \cite{MoTe2SDH}. To understand this discrepancy, we compare the intensity along q\textsubscript{$\perp$} (see SI Fig.4) with simulations and observation for various polytypes of NbSe\textsubscript{2} by Toshihiro Shimada et al.\cite{GIXRD_2} 2\textit{H}-MoTe\textsubscript{2} and 2\textit{H}-NbSe\textsubscript{2} share identical in-plane crystal structures and hence, would give similar intensity profiles along q\textsubscript{$\perp$} for various stacking orders (polytypes). Our q\textsubscript{$\perp$} (see SI Fig.3) scan closely matches the one reported by Toshihiro Shimada et al.\cite{GIXRD_2} on Se-GaAs, which is explained as a combination of 2\textit{Hb} and 3\textit{R} NbSe\textsubscript{2}. 2\textit{Hb} and 3\textit{R} both have the trigonal prismatic monolayer but the stacking sequence is different, with 3\textit{R} having a 3-layer periodicity as compared to 2-layer for 2\textit{Hb}. The value for MBE 2\textit{H}-MoTe\textsubscript{2} obtained by TEM is spatially local but X-ray beam for GI-XRD has a foot print of $\sim$2 mmx10 mm. The excess Te measured using XPS and the presence of stacking faults resulting in mixture of 2\textit{Hb} and 3\textit{R} phases could be the reason for the observed larger a and c lattice constants in MBE 2\textit{H}-MoTe\textsubscript{2} compared to bulk 2\textit{H}-MoTe\textsubscript{2}. 

SI Fig.3(a) shows the cs-TEM of the mixed phase sample B. For 1\textit{T'}- MoTe\textsubscript{2} is known to crystallize in P2\textsubscript{1}/m space group with lattice constants of a=6.33 {\AA}, b=3.48 {\AA} and c=13.82 {\AA}\cite{MoTe2SDH}, where b is very close to the lattice constant of  2\textit{H}-MoTe\textsubscript{2}. Therefore, for the sample C since the RHEED shows polycrystalline rings, irrespective of the in-plane rotational orientation of the film, we would expect to observe a ring corresponding to the $\{$010$\}$ set of planes in a similar scan as for Sample A (Fig.\ref{GIXRD}(a)). SI Fig.3(c) shows the GI-XRD on sample C ie. 1\textit{T'}-MoTe\textsubscript{2}. SI Fig.3(c) shows that there is no signal observed corresponding to 1\textit{T'}-MoTe\textsubscript{2}, the only peak is corresponding to CaF\textsubscript{2}. This is likely due to extremely low signal from the polycrystalline 1\textit{T'}-MoTe\textsubscript{2} thin film. It also shows the variability in crystallinity of CaF\textsubscript{2} from substrate to substrate, motivating the use of epi-ready GaAs substrates in Section B.

\subsubsection{Electrical measurements}

Sheet resistivity of the sample A and sample C were measured to be 5468 $\Omega$/$\square$ and 13255 $\Omega$/$\square$, respectively. This is of interest because film on sample A is 2\textit{H}-MoTe\textsubscript{2} and film on sample C is assigned to 1\textit{T'}-MoTe\textsubscript{2}. Traditionally 1\textit{T'}-MoTe\textsubscript{2} should be metallic and have lower resistance than 2\textit{H}-MoTe\textsubscript{2} but here the extensive oxidation of Sample C could be the cause of the significant increase in resistivity. TeO\textsubscript{2} glasses show semiconducting behavior\cite{TeO2}.

\section{Section B : M\MakeLowercase{o}T\MakeLowercase{e}\textsubscript{2} \MakeLowercase{on} G\MakeLowercase{a}A\MakeLowercase{s} (111)B}

In this section, 2 samples (sample D and sample E) of MoTe\textsubscript{2} on GaAs (111)B are discussed. post-growth anneal was done on MoTe\textsubscript{2} films on GaAs. 

\subsection{Growth conditions for sample D}

Sample D was grown at a growth temperature of 340$\degree$C and Te flux of 6.9x10\textsuperscript{-6} Torr (slightly higher than Sample A due to higher thermal conductivity of GaAs than CaF\textsubscript{2}). SI Fig.5(a), shows the growth sequence. The key step in this growth is the anneal of GaAs under Te prior to MoTe\textsubscript{2} growth to achieve smoother Te terminated surface. This is consistent with our previous report of Te anneal of GaAs\cite{vishJMR} prior to TMD superlattice growth. The change in growth rate between samples (see Table.1) is due to variability of Mo flux at the same e-beam power.

\subsection{Results and Discussion}
\subsubsection{RHEED}

For sample D, a pair of faint RHEED streaks with a spacing less than MoTe\textsubscript{2} were observed. Te has a hexagonal crystal structure with lattice constants of a=4.456 {\AA} and c=5.921 {\AA}\cite{TeLatConst}. Since the inplane lattice constant of Te is greater than that of 2\textit{H}-MoTe\textsubscript{2}, it was the initial suspect. With the aim to remove any excess Te in the film, a post-growth anneal at 380$\degree$C without any Te flux was done. But, as seen in SI Fig.5(b), the anneal doesn't remove this second set of streaks. Further analysis reveals that the ratio of spacing of the RHEED streaks from MoTe\textsubscript{2} and the newly observed streaks is $\sim$2. If the lines were from MoTe\textsubscript{2} and Te, the expected ratio is $\sim$1.2. So, its likely not due to elemental tellurium at the surface. One hypothesis is presence of ordered defects, which could be Te interstitials. Fig.\ref{GaAs}(a) shows RHEED intensity oscillations of the RHEED spectral point during the growth of MBE 2\textit{H}-MoTe\textsubscript{2}. This shows close to layer by layer growth. Simulation in SI Fig.7 shows the crest and the trough do not necessarily indicate a complete monolayer, variation in smoothness can cause shifts. But approximately, the period between crests corresponds to a monolayer. Increasing roughness or waviness in the film is the likely cause for decay with RHEED oscillation intensity in Fig.\ref{GaAs}(a). 

\subsubsection{Surface morphology, XRD and TEM}

It is very interesting to note that after cooling to room temperature, sample D shows Te crystallites on the surface. From AFM image (Fig.\ref{GaAs}(c)) and SEM image (SI Fig.6(a)) we can observe that these crystallites have preferential crystallographic orientation with the underlying GaAs  (111) with triangular symmetry. Also, we find that the height of these crystallites is $\sim$9 nm and about $\sim$50 nm wide. The fact that these crystallites are purely tellurium is confirmed by the markedly distinct lattice spacing compared to MoTe\textsubscript{2} as seen in the high resolution transmission electron microscopy (HRTEM) image shown in Fig.\ref{GaAs}(d). This is further confirmed by TEM Energy-dispersive X-ray spectroscopy (not shown). The HRTEM image (Fig.\ref{GaAs}(d)) also shows a high quality of MoTe\textsubscript{2} with a c-axis lattice spacing of 13.9 {\AA} consistent with 2\textit{H}-MoTe\textsubscript{2}. These crystallites could have likely been formed during the cooling process to room temperature. We don't observe Te crystallite formation on films on CaF\textsubscript{2} (see SI Fig.2). XRD scan in SI Fig.6((b) shows that at room temperature the (004) peak for sample D appears at 24.46$\degree$, which is lower than that for bulk 2\textit{H}-MoTe\textsubscript{2} and 1\textit{T'}-MoTe\textsubscript{2}\cite{MoTe2SDH}.  It corresponds to a c-spacing of 14.52 $\pm$ 0.05{\AA}. It is analyzed below along with (004) peak of sample E (see Table.4).

\subsubsection{Raman and electrical characterization}

\begin{figure}
	\includegraphics[width=0.8\columnwidth]{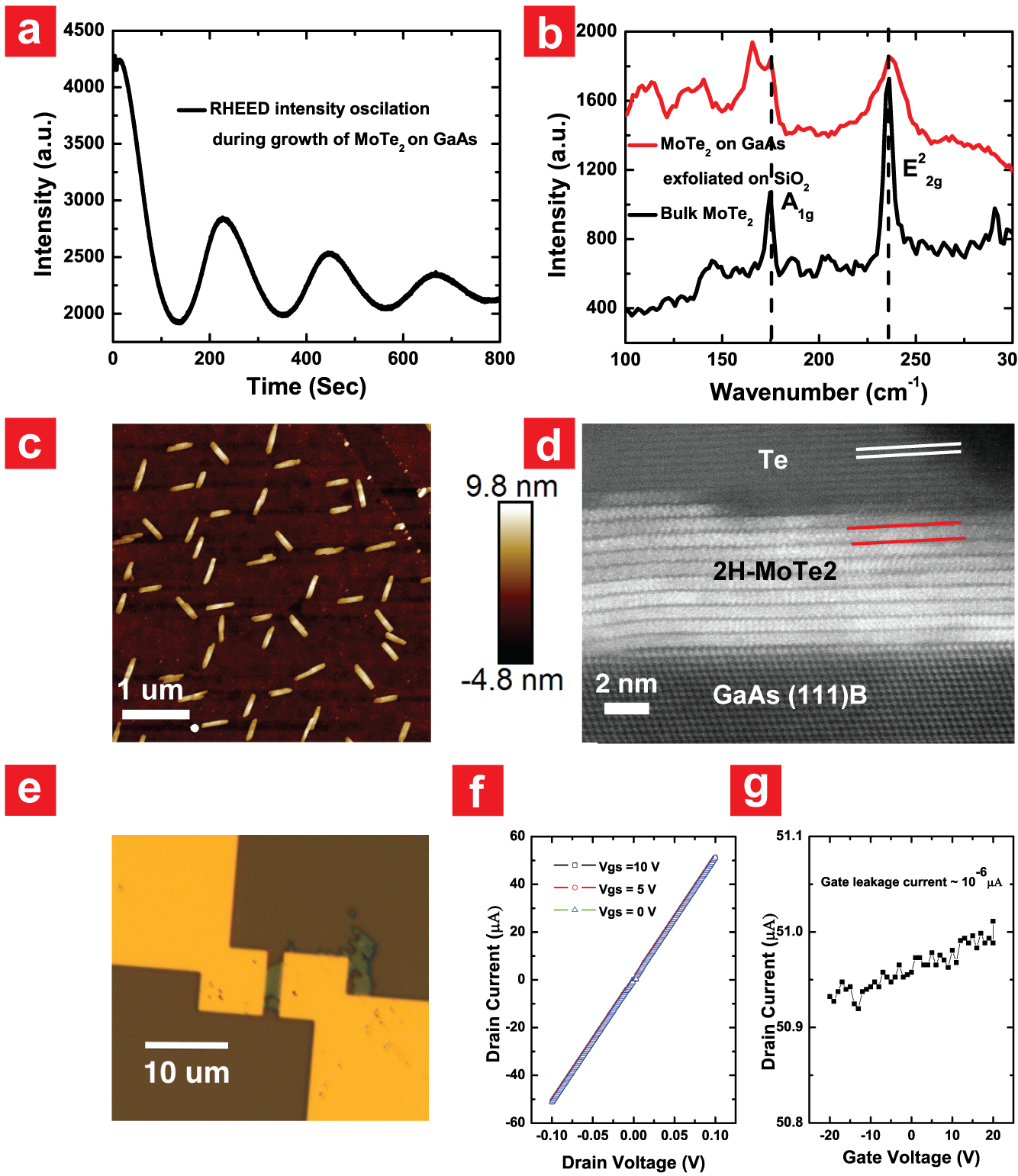}
	\caption{Sample D: (a) RHEED oscillations at the spectral point during growth of MoTe\textsubscript{2} shows a close to layer by layer growth with a period of $\sim$218 seconds per monolayer. (b) Raman from MoTe\textsubscript{2} grown on GaAs post-exfoliation, using scotch tape, on to SiO\textsubscript{2}/Si for a better signal. (c) The surface of the thin film post-growth shows several tellurium crystallites, surprisingly with preferential direction of orientation. (d) cross-sectional TEM shows the abrupt interface between GaAs and MoTe\textsubscript{2}, better quality of $\sim$9 monolayer MoTe\textsubscript{2} than on CaF\textsubscript{2} and pure tellurium crystallite with a significantly different lattice constant and contrast. The pair of white lines and the pair of red lines are a guide to the eye marking the difference in lattice constant of Te and 2\textit{H}-MoTe\textsubscript{2}. (e) Optical image of the contacts fabricated to measure transistor characteristics of as grown MoTe\textsubscript{2}. (f) shows the variation of drain current vs drain voltage at various gate biases and (g) shows that there is no gate modulation and the behavior is similar to a metal. \label{GaAs}}
\end{figure}

Since, the Raman signal from MoTe\textsubscript{2} on GaAs is quite weak\cite{vishJMR}, the as grown film was exfoliated and transferred to SiO\textsubscript{2}/Si substrate using a scotch tape. The transfer was performed to enhance the raman signal from the MoTe\textsubscript{2} due to cavity effect from SiO\textsubscript{2} as well as to eliminate the interference from the LO phonon raman peaks from GaAs. In Fig.\ref{GaAs}(b), the peaks below 150 cm\textsuperscript{-1} can be attributed to Te\cite{TeRaman}. E\textsubscript{2g} peak from transferred MoTe\textsubscript{2} from sample D is almost symmetric and peak position is consistent with that from bulk MoTe\textsubscript{2} at 235.8 cm\textsuperscript{-1}. The reason for broadening in the A\textsubscript{1g} peak is unclear. In order to test the electrical characteristic of the transferred MBE grown MoTe\textsubscript{2} film, contacts were made to the transferred flake. Using backgating, no modulation was observed (see Fig.\ref{GaAs}(e and f)).

\subsection{Growth conditions for sample E}

Sample E was grown on n\textsuperscript{$+$} GaAs for characterizations requiring conducting substrate. Prior to the post-growth anneal, the growth sequence for sample E was identical to sample D. The post-growth anneal for sample E  was done at 450$\degree$C for 3 mins and 550$\degree$C for 7 mins under Tellurium flux.  After this post-growth anneal, the sample was cooled under Te till growth temperature of 340$\degree$C and then capped with $\sim$100 nm Se during cool-down to room temperature for surface protection during sample transfer to other characterization tools. 

\subsection{Results and Discussion}

\subsubsection{LEED and XRD data}

From the phase diagram\cite{ASM}, one might expect mixed phase formation during the post-growth anneal at 550$\degree$C under Te. LEED and XRD was done to check the phase of the film grown. Prior to LEED, the Se cap was desorbed by annealing the sample in UHV at 300$\degree$C for $\sim$30 minutes. A LEED pattern measured with 40 eV electrons (Fig.\ref{LEED}(b) shows two sets of spots with the outer hexagonal pattern corresponding to the lattice constant of 2\textit{H}-MoTe\textsubscript{2} (`a' lattice constant using LEED = 3.57 $\pm$ 0.03 {\AA})and the inner pattern corresponding to an effective 2x2 superstructure. This is very interesting because it is consistent with the second set of RHEED streaks seen for sample D in SI Fig.5(b), discussed previously. One possible explanation for the extra spots is a change in the surface periodicity relative to the bulk 2\textit{H}-MoTe2 crystal structure due to a reconstruction or the presence of ordered defects. An alternate possibility is that the large electron spot size ($\sim$1 mm) may be averaging over three domains of 1\textit{T'}-MoTe\textsubscript{2} rotated by 60 degrees and 120 degrees (See SI Fig.8). 

To distinguish between these possibilities, temperature dependent XRD was done on sample E. At room temperature (002), (004), (006) and (008) peak 2$\theta$ positions of the as grown film are 12.17$\degree$, 25.02$\degree$, 38.02$\degree$ and 51.38$\degree$ respectively. The reported room temperature 2$\theta$ positions for the (002), (004), (006) and (008) peaks for 2\textit{H}-MoTe\textsubscript{2} are 12.66$\degree$, 25.48$\degree$, 38.63$\degree$ and 52.34$\degree$ respectively. D.H Keum et al.\cite{MoTe2SDH} report using temperature dependent XRD that the (004) 2$\theta$ peak of 2\textit{H}-MoTe\textsubscript{2} is at 25.5$\degree$ at room temperature and it shifts to slightly greater than 26$\degree$ at temperatures above 600$\degree$C corresponding to (004) 2$\theta$ peak of 1\textit{T'}-MoTe\textsubscript{2}. In our case, the (004) 2$\theta$ peak position is much lower than both peaks ($\sim$0.5$\degree$ lower than 2\textit{H}-MoTe\textsubscript{2}), which corresponds to a c-spacing of  14.25 $\pm$ 0.04{\AA}, and the peaks from the film is lost above 400$\degree$C. Therefore, without a chalcogen over pressure, MBE grown MoTe\textsubscript{2} dissociates between 400$\degree$C and 500$\degree$C. Also, sample D (see SI Fig.6(b)) which was grown without anneal at 550$\degree$C anneal has $\sim$0.5$\degree$ lower than sample E. Peak at smaller 2$\theta$ implies larger lattice constant but its origin is unclear yet. One possible explanation is the presence of excess tellurium in the crystal, which has been previously reported for bulk crystals\cite{MoSe2XRD}. The MoTe\textsubscript{2} phase diagram\cite{ASM} shows that 2\textit{H}-MoTe\textsubscript{2} is not a line compound. The 2\textit{H} phase of MoTe\textsubscript{2} can be formed inspite of a 1\% sub- or super- stoichiometric incorporation of tellurium. Hence, as XRD for both samples on GaAs does not show detectable peaks from the 1\textit{T'}-MoTe\textsubscript{2} phase, the 2x2 superstructure observed in LEED, is likely a surface feature rather than the presence of rotated domains of 1\textit{T'}-MoTe\textsubscript{2}. Lattice constants obtained from the various techniques described above have been tabulated in Table.4.

\begin{table}[]
\centering
\caption{In-plane and out of plane lattice constants for the 2\textit{H}-MoTe\textsubscript{2} samples in this study obtained by various techniques}
\label{my-label}
\begin{tabular}{|c|c|c|c|c|c|}
\hline
\begin{tabular}[c]{@{}c@{}}Sample\\ ID\end{tabular} & RHEED ({\AA}) & XRD ({\AA}) & GI-XRD ({\AA}) & LEED ({\AA}) & TEM ({\AA}) \\ \hline
A         & \begin{tabular}[c]{@{}c@{}}a = 3.5\\ $\pm$ 0.1\end{tabular}      &     & \begin{tabular}[c]{@{}l@{}}a = 3.64\\ $\pm$ 0.03\\ c = 14.4\\ $\pm$ 0.03\end{tabular}       &      & \begin{tabular}[c]{@{}c@{}}c = 13.9\\ $\pm$ 0.1\end{tabular}     \\ \hline
D         & \begin{tabular}[c]{@{}c@{}}a = 3.6\\ $\pm$ 0.05\end{tabular} & \begin{tabular}[c]{@{}c@{}}c = 14.52\\ $\pm$ 0.05\end{tabular}           &        &      & \begin{tabular}[c]{@{}c@{}}c = 13.8\\ $\pm$ 0.1\end{tabular}     \\ \hline
E         & \begin{tabular}[c]{@{}c@{}}a = 3.5\\ $\pm$ 0.1\end{tabular}  & \begin{tabular}[c]{@{}c@{}}c = 14.25\\ $\pm$ 0.04\end{tabular}          &        & \begin{tabular}[c]{@{}c@{}}a = 3.57\\ $\pm$ 0.03\end{tabular}      &     \\ \hline
Bulk 2\textit{H}        &  & \begin{tabular}[c]{@{}l@{}}a = 3.52085\cite{MoTe2SDH}\\ c = 13.9664\cite{MoTe2SDH}\end{tabular}           &        &      &     \\ \hline
Bulk 1\textit{T'}        &  & \begin{tabular}[c]{@{}l@{}}a = 6.3274\cite{MoTe2SDH}\\ b = 3.4755\cite{MoTe2SDH}\\ c = 13.8100\cite{MoTe2SDH}\end{tabular}           &        &      &     \\ \hline
\end{tabular}
\end{table}

\subsubsection{Sensitivity to air exposure of MoTe\textsubscript{2}}

\begin{figure}
	\includegraphics[width=\columnwidth]{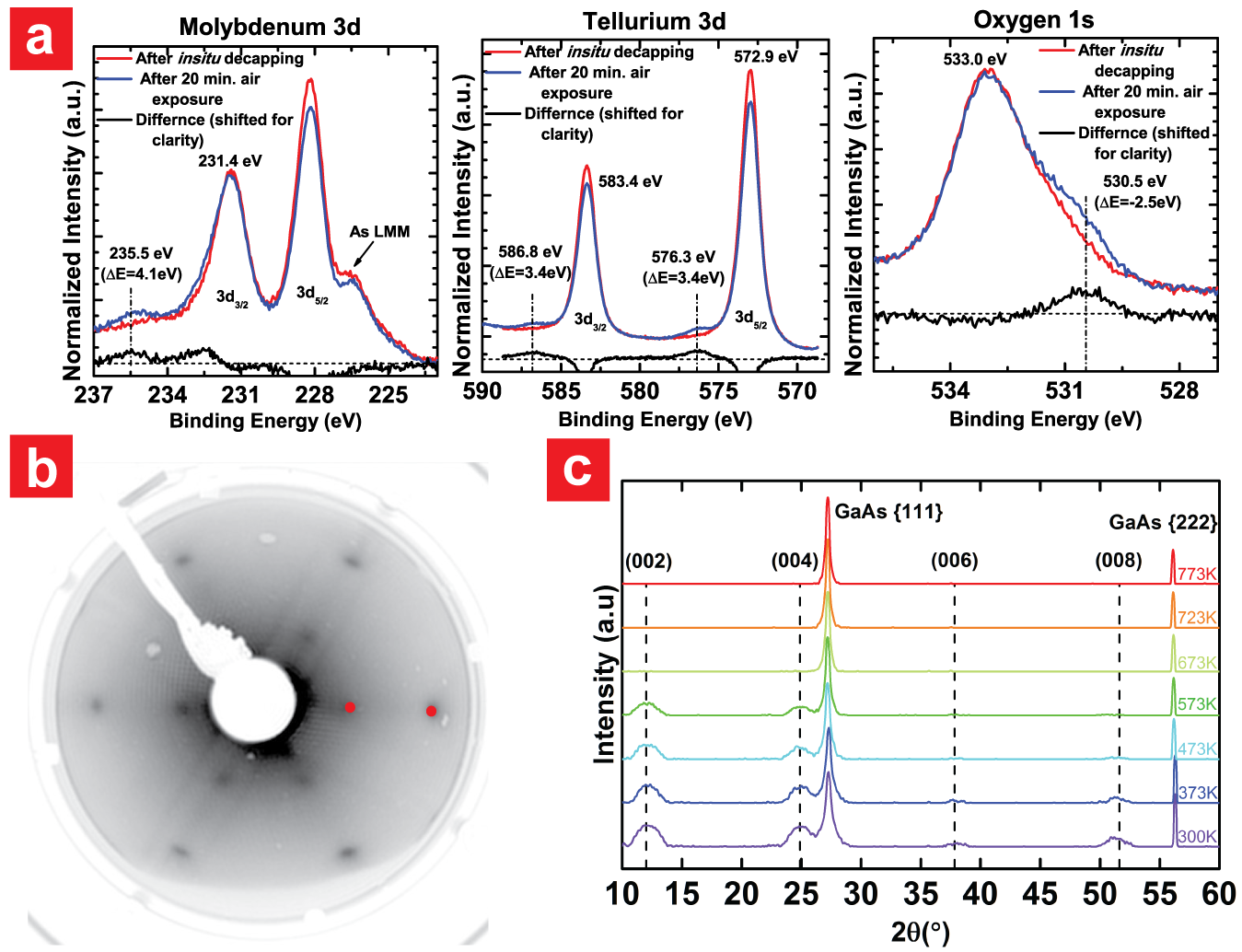}
	\caption{ (a) XPS on sample E (MoTe\textsubscript{2} on GaAs (111)B) post Se decapping in UHV, prior to and post-air exposure. The dash-dot lines in all 3 spectra show the emergence of the oxide peak (b) LEED from sample E post-decapping without any air exposure, the red spots are guide to the eye demonstrating the 2x2 superstructure (c) Temperature dependent XRD on sample E under nitrogen environment showing the phase of MoTe\textsubscript{2} and its thermal stability\label{LEED}}
\end{figure}

To understand the ease of oxidation of MoTe\textsubscript{2}, on sample E, the Se cap was removed from Sample E by heating in a UHV system, followed by in situ XPS. After the initial XPS measurements, the sample was exposed to air for 20 mins and then XPS done again. It can be clearly seen from Fig.\ref{LEED}(a) that oxide peaks appear in both Mo and Te XPS spectrum. Approximately, 8$\%$ of the surface area under goes oxidation in 20 mins. Therefore, through the process of exfoliation and fabrication the film is likely to undergo extensive oxidation. For TeO\textsubscript{2} glasses the conductivity goes down with lowering temperature but the conduction mechanism is by charge hopping\cite{TeO2}. The Se decapping is done in an oxide MBE system that is connected to LEED and XPS system to avoid air exposure.The O 1s signal observed in XPS (see Fig.\ref{LEED}(a)) prior is air exposure is likely due to physisorbed oxygen, post-decapping, from the oxide MBE chamber (base pressure of $\sim$1x10\textsuperscript{-8} Torr). This is consistent with the fact that electronegativity difference between Mo and Te is 0.3 eV but that between O and Te is 1.4 eV, making the compound prone to oxidation\cite{XPSpaper,bernedeXPS}. Effect of air exposure in MBE grown films could be exacerbated by oxidation at the edges of MoTe\textsubscript{2} grains, similar to that reported for MBE grown WSe\textsubscript{2} films\cite{VishWSe2air}.  

\section{Conclusion}
This work employs extensive large area structural and chemical characterization of MBE grown few layer MoTe\textsubscript{2}, with complementing electrical characterization. In section A, we show that for growth of few layer 2\textit{H}-MoTe\textsubscript{2} at a low temperature of 340$\degree$C and growth rate of $\sim$6 mins/ML, we need an incident Te:Mo flux greater than 100. The 2\textit{H} and 1\textit{T'} phases formed in different samples have distinct signatures in RHEED, Raman and XPS, but the Te:Mo stoichiometry determined by XPS is greater than 2 for both. GI-XRD shows a small grain size of $\sim$90 {\AA}, twinning and a higher-than-expected c-spacing in 2\textit{H}-MoTe\textsubscript{2} on CaF\textsubscript{2}. XRD on MBE 2\textit{H}-MoTe\textsubscript{2} on GaAs in section B also shows larger c spacing than both bulk 2\textit{H}-MoTe\textsubscript{2} and bulk 1\textit{T'}-MoTe\textsubscript{2}. Growth on GaAs shows Te crystallite formation on the surface, and a 2x2 pattern in RHEED and LEED. All these have been hypothesized as signs of excess Te incorporation into 2\textit{H}-MoTe\textsubscript{2} during growth. At ambient pressure in N\textsubscript{2} atmosphere, MBE 2\textit{H}-MoTe\textsubscript{2} on GaAs is only stable up to 300$\degree$C. Excess Te in the film can explain the high electrical conductivity, non-modulating behaviour and easy dissociation of the film with increasing temperature prior to phase transition to the 1\textit{T'} phase, a more stable phase at higher temperatures. Finally, we demonstrate the swift oxidation ($\sim$8$\%$ surface area in 20 mins) of the MBE MoTe\textsubscript{2} film on exposure to air.  With the various complementing large area and local characterizations, this study has provided insight into the few layer MBE growth of Mo-Te system on 3D substrates.

\section{Experimental Methods}

\subsubsection{Raman Spectroscopy}

Raman measurements were performed in the backscattering configuration using a WITec Alpha 300 system at room temperature. Measurement was done using a 100x objective, 1800 grooves/mm grating, 488 nm laser and 0.75 mW power.

\subsubsection{X-ray Photoelectron Spectroscopy}

XPS on the CaF\textsubscript{2} (Series A) samples was carried out ex-situ using a monochromated Al K$\alpha$ source (h$\nu$ = 1486.7 eV) and an Omicron Argus detector (MCD-128) operating with a pass energy of 15 eV. XPS spectra were acquired at a pass energy of 15 eV and take-off angle (defined with respect to the sample surface) of 45$\degree$ and 80$\degree$.  For XPS peak deconvolution, the spectral analysis software Aanalyzer was employed, where Voigt line shapes and an active Shirley background were used for peak fitting\cite{herrera2002chemical}.

XPS on the 2\textit{H}-MoTe\textsubscript{2}/n\textsuperscript{$+$}GaAs (Series B) samples was measured using a non-monochromated Al K$\alpha$ source and a Scienta R4000 electron analyzer operating at a pass energy of 100 eV. All spectra were measured at normal emission, i.e. 90 degrees relative to the sample surface. Central peak locations were determined by Lorentzian fits with linear backgrounds.

\subsubsection{X-ray Diffraction}

Out of plane XRD and temperature dependent XRD on the 2\textit{H}-MoTe\textsubscript{2}/GaAs samples is done using the Rigaku SmartLab X-Ray Diffractometer with Cu K$\alpha$ X-ray source. The GI-XRD is done using the G2 hutch at the CHESS beamline (http://www.chess.cornell.edu/gline/G2.htm), operating with a x-ray energy of 11.31 KeV.

\subsubsection{Transmission Electron Microscopy}

The atomic structure analysis for sample A and sample D was carried out on FEI Titan 80-300 Transmission Electron Microscope operated at 300 kV. 

TEM on sample B was done using JOEL ARM200F atomic resolution analytical microscope.

\subsubsection{LEED}

LEED on the 2\textit{H}-MoTe\textsubscript{2}/n\textsuperscript{$+$}GaAs samples was measured using a Specs ErLEED 3000 system with an incident electron energy of 40 eV. The electron spot size was approximately 1 mm in diameter, and the total angular field of view was 100 degrees. Following the Se decapping at 300$\degree$C, in situ LEED and XPS measurements were both performed at room temperature in an analysis chamber with pressure below $\sim$1x10\textsuperscript{-10} Torr.

\section{Acknowledgments}
This work was supported in part by the Center for Low Energy Systems Technology (LEAST), one of six centers of STARnet, a Semiconductor Research Corporation program sponsored by MARCO and DARPA. This work made use of the Cornell Center for Materials Research Shared Facilities which are supported through the NSF MRSEC program (DMR-1120296). This work is based upon research conducted at the Cornell High Energy Synchrotron Source (CHESS) which is supported by the National Science Foundation and the National Institutes of Health/National Institute of General Medical Sciences under NSF award DMR-1332208. MBE growth is in part supported by NSF Grant DMR 1400432 and NSF-EFRI 2DARE Grant DMR 1433490.

\section{Supplementary Information}

\begin{figure}[!h]
	\includegraphics[width=0.5\columnwidth]{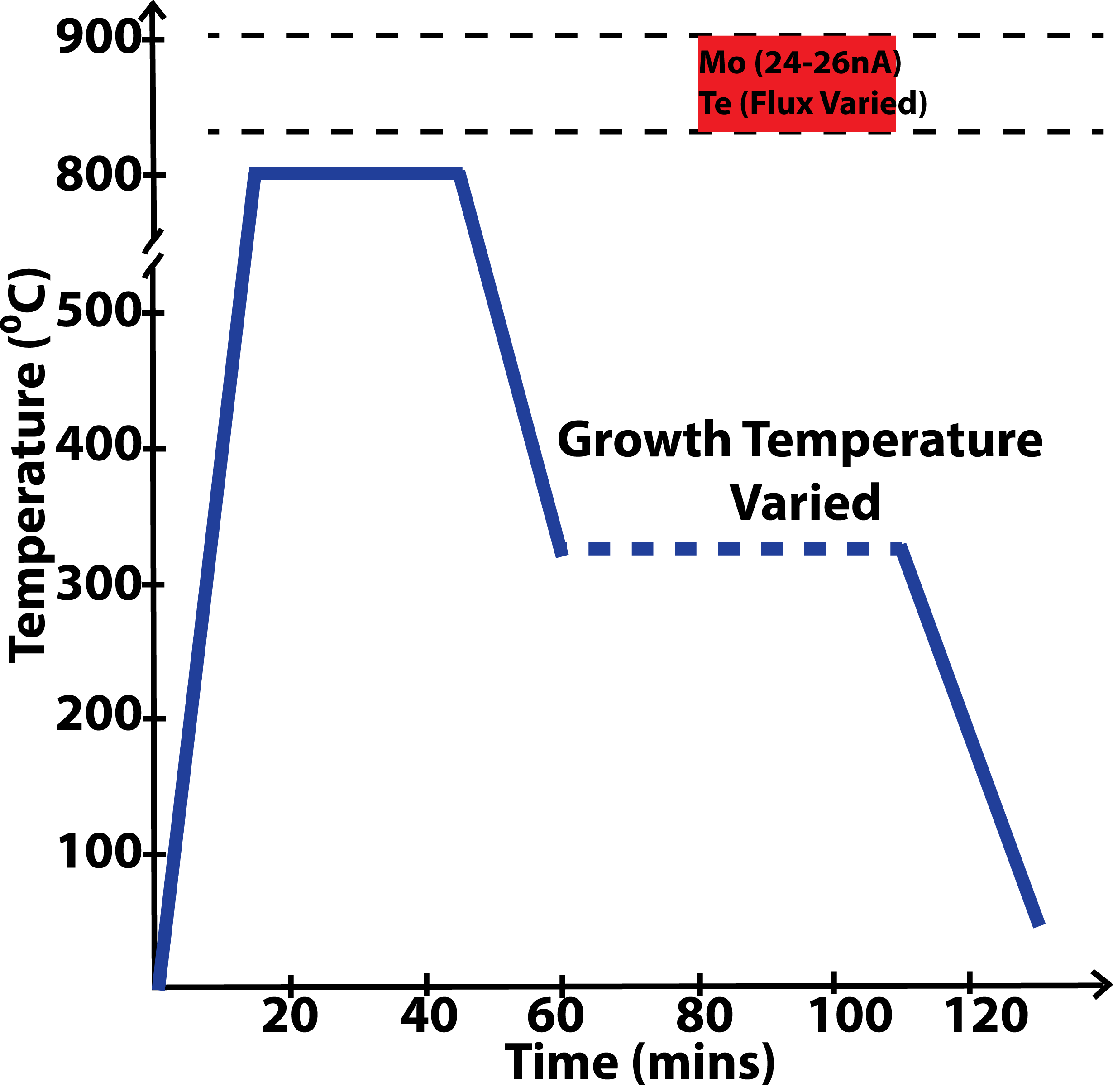}
	\caption{ General schematic of the growth diagram of MoTe\textsubscript{2} thin film on CaF\textsubscript{2}.\label{SIone}}
\end{figure}

\begin{figure}
	\includegraphics[width=\columnwidth]{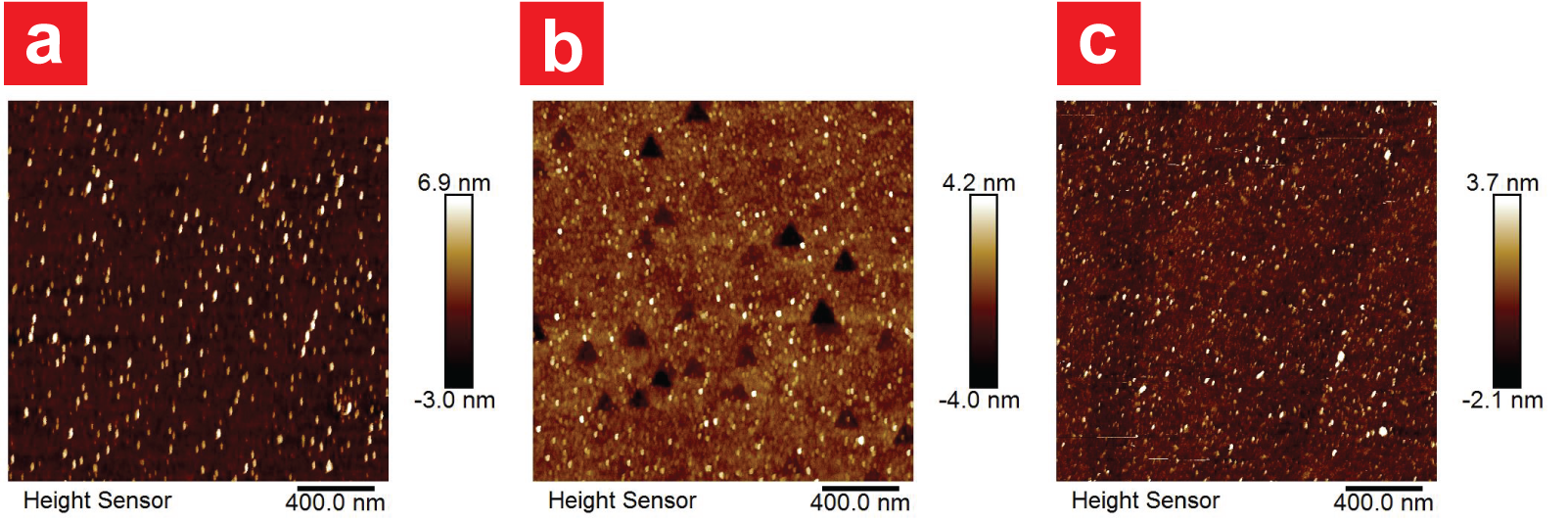}
	\caption{ AFM of (a) sample A, (b) sample B and (c) sample C.\label{SItwo}}
\end{figure}

\begin{figure}
	\includegraphics[width=\columnwidth]{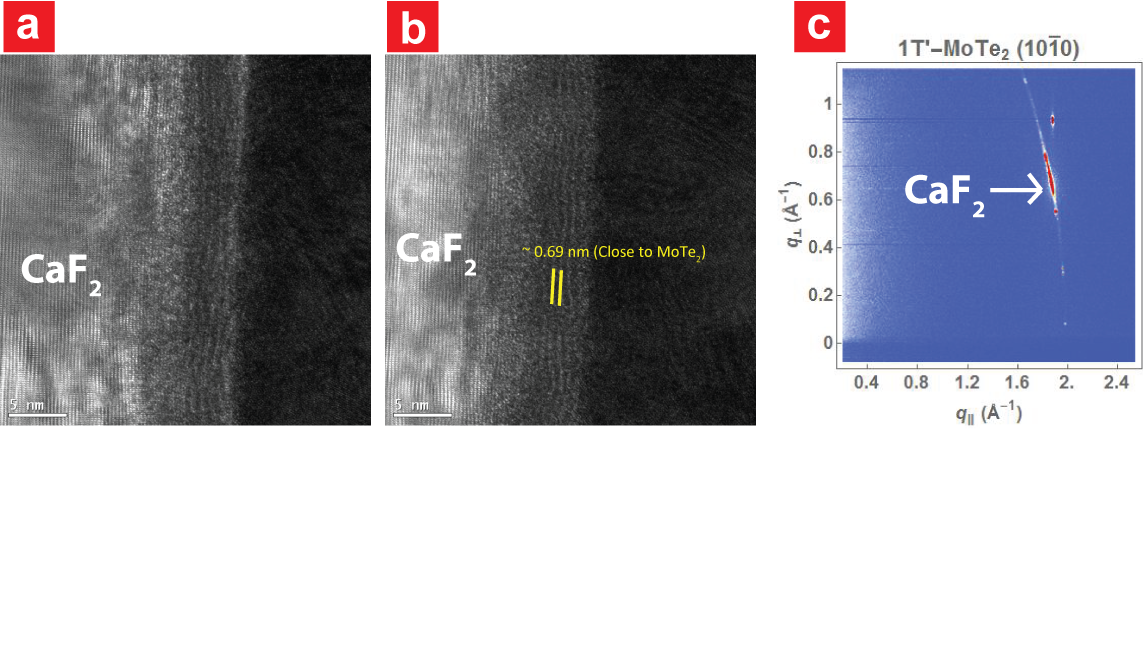}
	\caption{ (a and b)cs-HRTEM of sample B that is the mixed phase MoTe\textsubscript{2} and (c) GI-XRD from Sample C that is 1T' MoTe\textsubscript{2}. Only signal from CaF\textsubscript{2} is observed, none from MoTe\textsubscript{2} thin film.\label{SIthree}}
\end{figure}

\begin{figure}
	\includegraphics[width=\columnwidth]{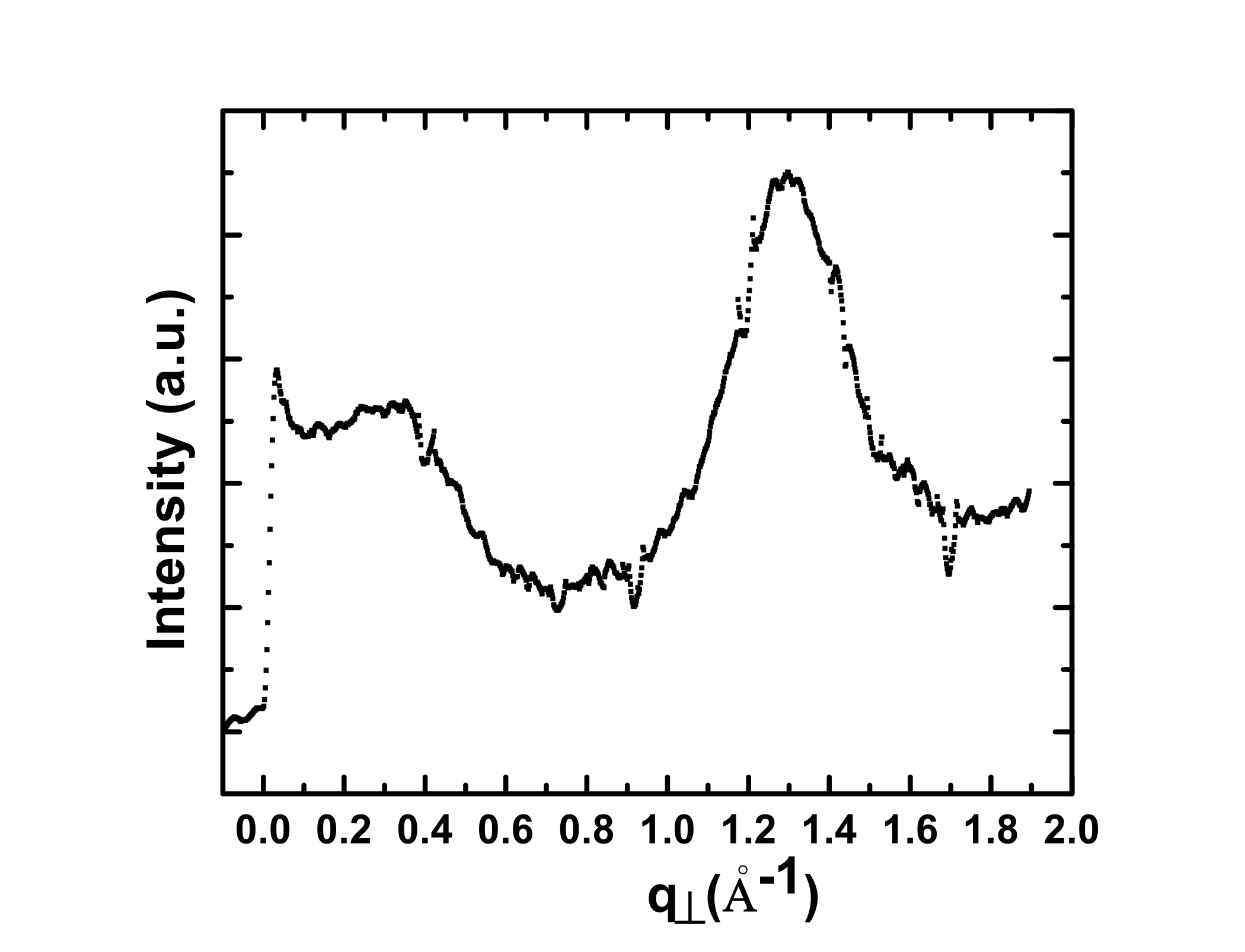}
	\caption{ Bragg rod profile for $\{$10$\bar{1}$x$\}$ peaks from 2\textit{H}-MoTe\textsubscript{2} /CaF\textsubscript{2} \label{SIfour}}
\end{figure}

\begin{figure}
	\includegraphics[width=\columnwidth]{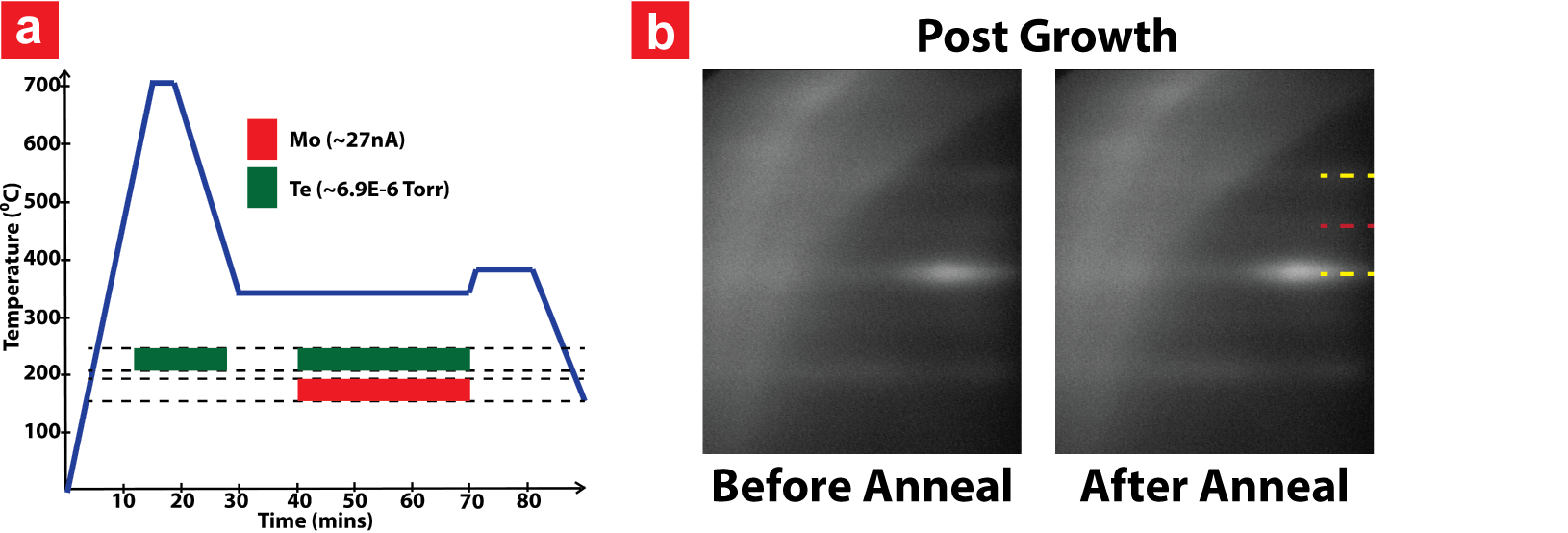}
	\caption{ (a) Growth sequence for Sample D and RHEED of MoTe\textsubscript{2} on GaAs post growth, (b) pre anneal and (b) post anneal. The dashed lines are guide to the eye, yellow for MoTe\textsubscript{2} and red for the additional set of lines observed \label{SIfive}}
\end{figure}

\begin{figure}
	\includegraphics[width=\columnwidth]{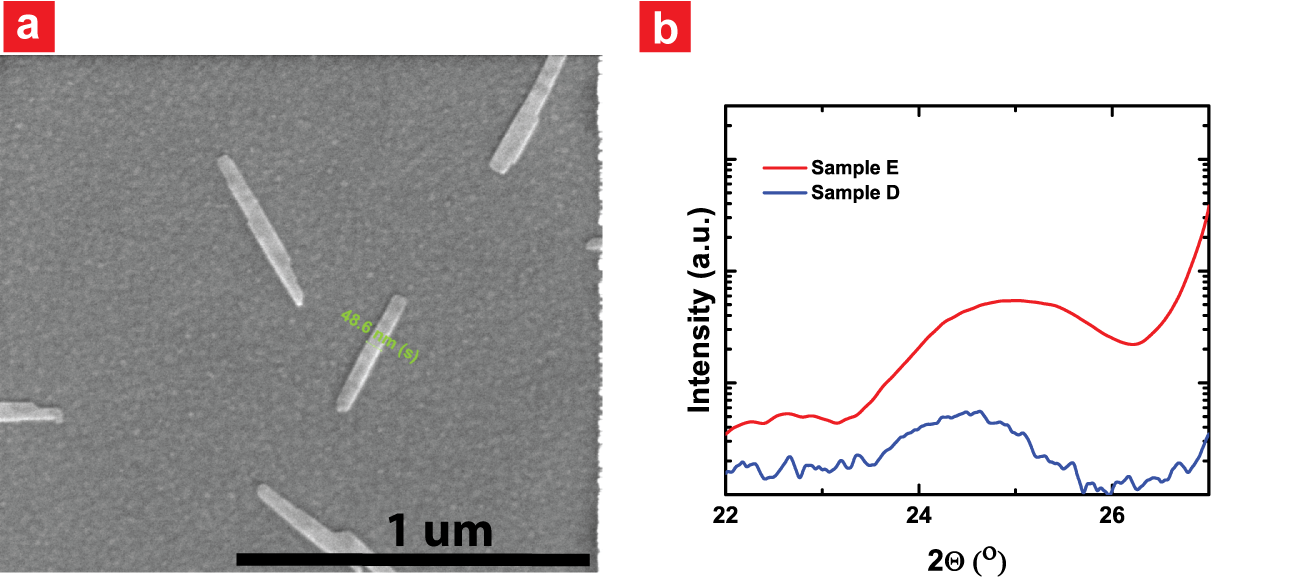} 
	\caption{ (a) SEM of sample D showing electrically isolated Tellurium crystallites and (b) the zoom in of the (004) XRD peak from sample D and sample E at room temperature.\label{SIsix}}
\end{figure}

\begin{figure}
	\includegraphics[width=\columnwidth]{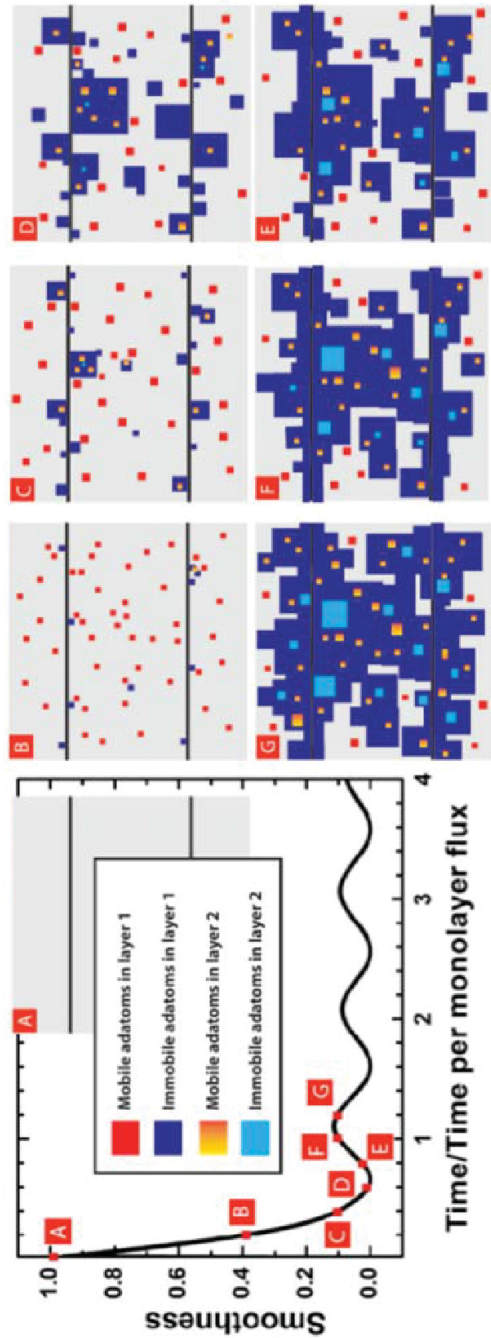}
	\caption{ Simulation of non-equilibrium surface morphology evolution following the subsection 6.3.3 of the book titled ``Materials Fundamentals of Molecular Beam Epitaxy" by J.Tsao. RHEED intensity follows closely surface roughness which is a consequence of different number of layer coverage spacially. The 2 black lines corresponds to atomic steps on the substrate where nucleation could occur. In this simulation all kinetic parameters are taken to be 20/time per monolayer flux, expect for the kinetic rate constant for attachment of adatoms to the lower step edge (``down'' step), which is slightly higher at 30/time per monolayer flux. \label{SIseven}}
\end{figure}

\clearpage
	
	\subsubsection{Possible origins of extra spots in low energy electron diffraction (LEED) patterns from bulk MoTe\textsubscript{2}}
Fig. 8a shows the crystal structure of a monolayer transition metal dichalcogenide with undistorted octahedral transition metal coordination. Lines of metal atoms run in three different directions (referred to as 1, 2, and 3; see inset of Fig. 8b), where each line is separated by a distance d. In the distorted structure, shown in Fig. 8b, the metal atoms shift, creating zigzag chains of metal atoms where two neighboring lines are separated by a smaller distance d-$\delta$. These chains run along a single direction (in this case, direction 1), causing the unit cell to change from hexagonal to rectangular symmetry. In reciprocal space, extra Bragg reflections appear corresponding to the change in symmetry (Fig. 8c and 8d). Zigzag chains along directions 1, 2, and 3 are degenerate in energy, so it is reasonable to expect domains of different orientation in a macroscopic material. In this case, if the domain size is smaller than the electron beam, a LEED experiment will measure spatially averaged patterns from all three directions (Fig. 8e). The resulting Bragg reflections are identical to those caused by a 2x2 superstructure of the original undistorted crystal.
Bulk 1\textit{T'}-MoTe\textsubscript{2}  is formed by stacking individual monolayers with distorted octahedral Mo coordination. At a given incident electron energy, the out-of-plane periodicity determines the intensity of LEED spots but not the position of Bragg reflections. So a bulk 1\textit{T'}-MoTe\textsubscript{2} sample with multiple domains would produce Bragg reflections as shown in Fig. 8e. Alternatively, a bulk 2\textit{H}-MoTe\textsubscript{2} sample would produce Bragg reflections as shown in Fig. 8c. (The 2\textit{H} structure has trigonal Mo coordination, but retains the same hexagonal symmetry assumed for Fig. 8c.) If the sample surface acquired a larger 2x2 periodicity through a surface reconstruction or the ordering of surface defects, the surface sensitivity of LEED would ensure that the resulting Bragg reflections would appear as in Fig. 8e. Because 2\textit{H}-MoTe\textsubscript{2} and the distorted octahedral structures exhibit the same in-plane lattice constant, the Bragg reflections measured by LEED would appear at identical locations to bulk 1\textit{T'}-MoTe\textsubscript{2} with multiple domains.

\begin{figure}
	\includegraphics[width=\columnwidth]{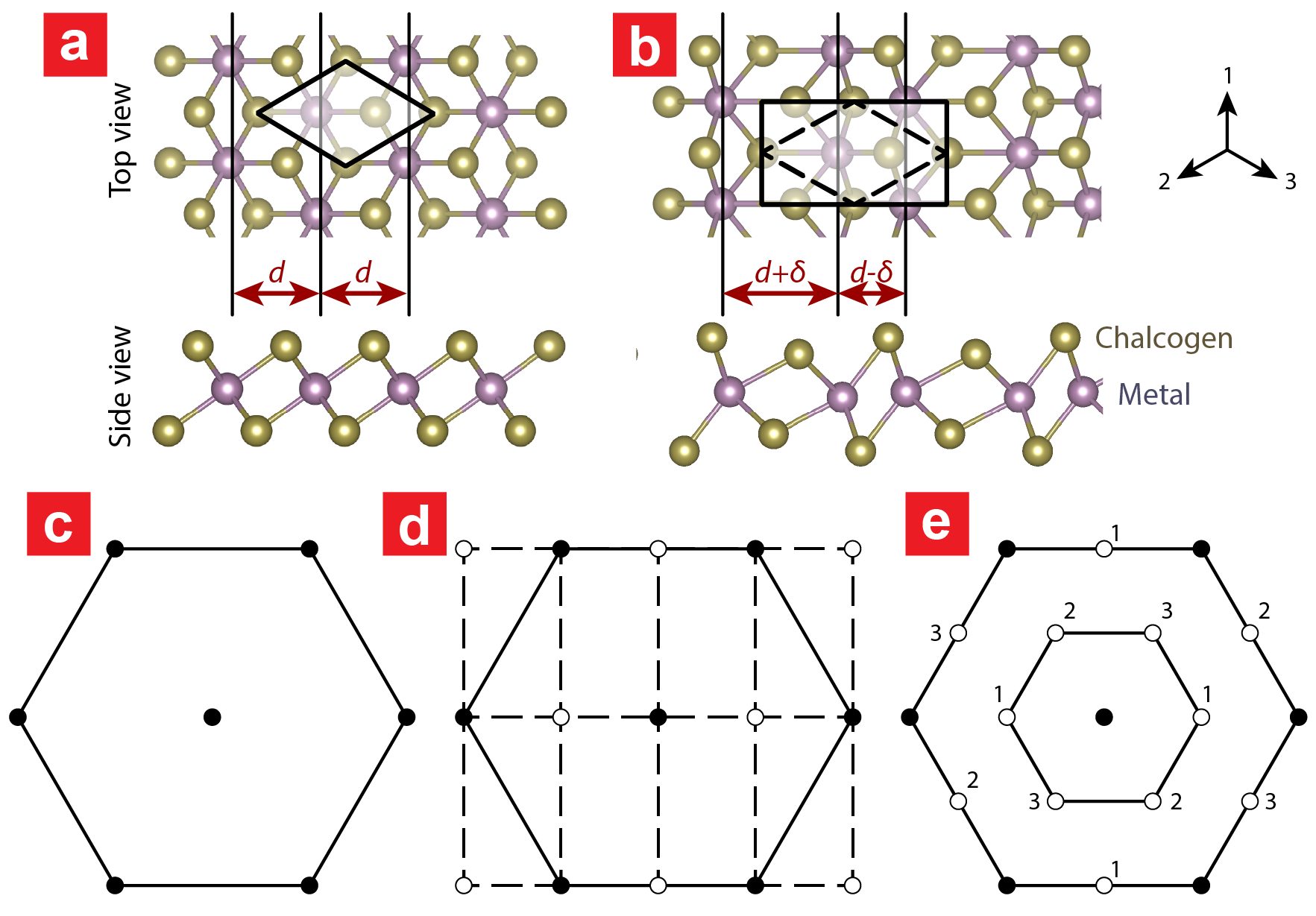}
	\caption{ (a), (b) Schematics showing crystal structures of monolayer transition metal dichalcogenides with undistorted and distorted octahedral transition metal coordination, respectively. (c), (d) Bragg reflections from the structures in (a) and (b), respectively, where filled (open) circles are from undistorted (distorted) structures. (e) Spatially averaged Bragg reflections resulting from distortions along three different directions, where the open circles are labeled according to the direction of the corresponding distortion.\label{SIeight}}
\end{figure}

\clearpage
\section{References}

\bibliography{MoTe2ref} {}

\end{document}